\documentclass[aps,prc,floatfix,twocolumn,nofootinbib,superscriptaddress]{revtex4-1}

\usepackage[]{graphicx}
\usepackage{color}

\usepackage{amsmath,amssymb,amsfonts}

\usepackage{slashed}

\begin{document}

\title{Gluon emission from heavy quarks in dense nuclear matter}

\author{Le Zhang}
\affiliation{College of Physics and Electronic Science, Hubei Normal University, Huangshi, 435002, China  }
\affiliation{Institute of Particle Physics and Key Laboratory of Quark and Lepton Physics (MOE), Central China Normal University, Wuhan, 430079, China  }

\author{De-Fu Hou}
\affiliation{Institute of Particle Physics and Key Laboratory of Quark and Lepton Physics (MOE), Central China Normal University, Wuhan, 430079, China  }

\author{Guang-You Qin}
\affiliation{Institute of Particle Physics and Key Laboratory of Quark and Lepton Physics (MOE), Central China Normal University, Wuhan, 430079, China  }

\date{\today}
\begin{abstract}

We study the medium-induced gluon emission process experienced by a hard jet parton propagating through the dense nuclear matter in the framework of deep inelastic scattering off a large nucleus.
We work beyond the collinear rescattering expansion and the soft-gluon emission limit, and derive a closed formula for the medium-induced single gluon emission spectrum from a heavy or light quark jet interacting with the dense nuclear medium via transverse and longitudinal scatterings.
Without performing the collinear rescattering expansion, the medium-induced gluon emission spectrum is controlled by the full distribution of the differential elastic scattering rates between the propagating partons and the medium constituents.
We further show that if one utilizes heavy static scattering centers for the traversed nuclear matter and takes the soft-gluon emission limit, our result can reduce to the first order in opacity Djordjevic-Gyulassy-Levai-Vitev formula.

\end{abstract}
\maketitle

\section{Introduction}

High-energy heavy-ion collisions at the Large Hadron Collider (LHC) and the Relativistic Heavy-Ion Collider (RHIC) can create the strongly-interacting quark-gluon plasma (QGP), a state of matter consisting of deconfined quarks and gluons.
Hard partonic jets, which are produced from early hard scatterings, provide useful tools to study such highly-excited QCD matter.
During their passage through the QGP matter, jets may interact with the medium constituents via elastic and inelastic collisions.
The interaction between jets and medium not only causes the energy loss of hard partons, but also change the internal structure of full jet shower.
Such phenomenon is generally referred to as jet quenching \cite{Wang:1991xy, Qin:2015srf, Blaizot:2015lma, Majumder:2010qh}.
There have been a wealth of experimental evidences and phenomenological studies on jet-medium interaction and jet quenching in the LHC and RHIC heavy-ion collision programs, such as the suppressions of high transverse momentum ($p_T$) hadron productions \cite{Abelev:2012hxa, Aad:2015wga, CMS:2012aa, Burke:2013yra, Xu:2014tda, Chien:2015vja, Andres:2016iys, Cao:2017hhk, Zigic:2018ovr} and jet productions \cite{Adam:2015ewa, Aad:2014bxa, Khachatryan:2016jfl, Qin:2010mn, Young:2011qx, Dai:2012am, Wang:2013cia, Blaizot:2013hx, Mehtar-Tani:2014yea, Cao:2017qpx, He:2018xjv}, and the nuclear modification of jet-related correlations \cite{Aad:2010bu, Chatrchyan:2012gt, Qin:2009bk, Chen:2016vem, Chen:2016cof, Chen:2017zte, Luo:2018pto, Zhang:2018urd, Kang:2018wrs}, jet shape and jet substructure observables \cite{Chatrchyan:2013kwa, Aad:2014wha, Chang:2016gjp, Casalderrey-Solana:2016jvj, Tachibana:2017syd, KunnawalkamElayavalli:2017hxo, Brewer:2017fqy, Chien:2018dfn}.

From the theoretical side, tremendous effort has been devoted to our understanding of parton energy loss and jet modification in both cold and hot dense nuclear matter.
For example, the effects of binary elastic collisions of hard partons with medium constituents have been studied in various literatures \cite{Bjorken:1982tu, Braaten:1991we, Djordjevic:2006tw, Qin:2007rn, Qin:2012fua}.
The medium-induced radiative process experienced by hard partons in dense nuclear matter have been investigated in details as well.
There currently exist a few theoretical schemes on radiative parton energy loss based on the perturbative QCD framework: Baier-Dokshitzer-Mueller-Peigne-Schiff-Zakharov (BDMPS-Z) \cite{Baier:1996kr, Baier:1996sk, Baier:1998kq, Zakharov:1996fv, Zakharov:1997uu}, Djordjevic-Gyulassy-Levai-Vitev (DGLV) \cite{Gyulassy:1999zd, Gyulassy:2000er, Djordjevic:2003zk, Blagojevic:2018nve, Sievert:2018imd},
Armesto-Salgado-Wiedemann (ASW) \cite{Wiedemann:2000za, Wiedemann:2000tf, Armesto:2003jh}, Arnold-Moore-Yaffe (AMY) \cite{Arnold:2001ba, Arnold:2002ja, CaronHuot:2010bp} and higher twist (HT) \cite{Guo:2000nz, Wang:2001ifa, Zhang:2003wk, Majumder:2009ge} formalisms.
On the other hand, much recent phenomenological effort (see e.g., Ref. \cite{Burke:2013yra}) has been devoted to quantitative extraction of various jet transport coefficients, such as $\hat{q}$ which quantifies the transverse momentum squared transferred per unit path length between hard jet partons and soft nuclear medium \cite{Baier:1996sk}.

While the study of jet quenching in heavy-ion collisions has already entered quantitative era, various systematic uncertainties still exist from the theory side, such as different implementations of jet-medium interaction effects and various approximations made in the foundations of jet quenching formalisms.
Regarding medium-induced radiative parton energy loss, some jet quenching formalisms (e.g., GLV and ASW) take the soft approximation for the emitted gluons assuming the emitted gluon energy is much smaller than the parent parton energy, while others (e.g., BDMPS-Z and HT) take collinear expansion for the rescatterings with the medium constituents assuming the exchanged transverse momentum between jet and medium is small and can be approximated by Gaussian distribution.
These issues have already been pointed out in Ref. \cite{Armesto:2011ht} in which a detailed comparison of different jet energy loss schemes is also provided.
Some recent studies have been performed on relaxing various theoretical approximations.
For example, Ref. \cite{Apolinario:2014csa} includes the non-eikonal corrections for medium-induced emission within the path integral formalism.
Refs. \cite{Zhang:2018kkn, ZhangYY} generalize the HT formalism beyond the collinear rescattering expansion; it is interesting that such generalized HT formalism reduces to the GLV formalism in the soft-gluon emission limit.
Refs. \cite{Blagojevic:2018nve, Sievert:2018imd} reinvestigate the GLV (DGLV) formalism by relaxing the soft-gluon emission approximation.

In this work, we study the medium-induced gluon emission from a massive quark jet which scatters off the medium constituents during its passage through the dense nuclear medium, within the framework of deep-inelastic scattering (DIS) off a large nucleus.
In particular, we work beyond the collinear rescattering expansion and soft-gluon emission approximation, and derive a closed formula for the medium-induced single gluon emission spectrum from a heavy quark jet interacting with medium constituents via both transverse and longitudinal scatterings.
Without performing the collinear expansion, our medium-induced gluon emission spectrum is controlled by the full distribution of the differential elastic scattering rates between the propagating partons and medium constituents.
As compared to our previous work~\cite{Zhang:2018kkn}, the current work focuses on the charged current interaction channel in DIS and study the medium-induced gluon emission for both massive and massless quarks on the same footing.
Thus it can be viewed as a generalization of the higher twist formalism \cite{Guo:2000nz, Wang:2001ifa, Majumder:2009ge, Zhang:2003wk} for both heavy and light flavor medium-induced radiative process.
We further show that if one utilizes heavy static scattering centers for the traversed dense nuclear matter and takes the soft-gluon emission limit, our result can reduce to the first order in opacity GLV and DGLV (with zero effective mass for radiated gluon) formulae on medium-induced gluon emission \cite{Gyulassy:2000er, Gyulassy:2000gk, Djordjevic:2003zk}.

The paper is organized as follows. In Sec.~II, we present the gluon emission from a heavy (or light) quark jet at leading twist in the DIS framework. In Sec.~III, we derive the medium-induced single gluon emission spectrum for a hard quark jet propagating through the dense nuclear matter. Some details of our main results are presented in Appendix A.
The last section contains our summary.

\section{gluon emission at leading twist}

Consider the process in which a  heavy or light quark jet is produced in the framework of semi-inclusive deep inelastic scattering (DIS) off a large nucleus,
\begin{eqnarray}
 l(L_1) + A(A p) \to \nu_l(L_2) + q(l_q) + X(P_X).
\end{eqnarray}
Here $L_1$ and $L_2$ represent the momenta of the incoming lepton and outgoing neutrino. $Ap$ is the momentum of the incoming nucleus with the nucleon number $A$, and $p = [p^+, p^-, \mathbf{p}_\perp] = [p^+, \frac{m_N}{2 p^+}, \mathbf{0}_\perp] \approx [p^+, 0, \mathbf{0}_\perp]$ is the momentum of each nucleon in the nucleus ($m_N$ is the mass of the nucleon and is neglected in the high energy limit). The outgoing quark jet has the mass $M$ and carries the momentum $l_q$.
In this work, we focus on the charged current interaction channel since it allows us to study the medium-induced gluon emission for heavy (massive) and light (massless) quarks together on the same footing.
In the charged current interaction channel, the four-momentum of the exchanged $W$ boson is: $q = L_2 - L_1 = [-x_B p^+, q^-, \mathbf{0}_\perp]$, with $x_B = Q^2 / (2p^+q^-)$ being the Bjorken fraction variable and $Q^2=-q^2$ being the invariant mass of the exchanged $W$ boson. Here we assume $Q^2\ll m_W^2$.

The differential cross section for the above semi-inclusive DIS process can be written as:
\begin{eqnarray}
E_{L_2}  \frac{ d\sigma_{DIS}}{d^3\mathbf{L}_2} = \frac{G^2_F}{(4\pi)^3 s}  L_{\mu\nu}   W^{\mu\nu}.
\end{eqnarray}
Here $G_F$ is the four fermion coupling and $s=(p+L_1)^2$ is the center-of-mass energy of the lepton-nucleon collision system.
The leptonic tensor $L_{\mu\nu}$ is given as:
\begin{eqnarray}
L_{\mu\nu} = \frac{1} {2} \rm Tr[\slashed{L}_1  (1+\gamma^5) \gamma_\mu \slashed{L}_2 \gamma_\nu (1-\gamma^5)].
\end{eqnarray}
The semi-inclusive hadronic tensor $W^{\mu\nu}$ is given as:
\begin{eqnarray}
W^{\mu\nu} &=&\frac{1}{2}\sum_{X} (2\pi)^4 \delta^4 (q + A p - P_X-l_q)
\nonumber\\
&\times &\langle A| J_\mu(0)|X\rangle \langle X|{J_\nu}^\dag(0)|A\rangle,
\end{eqnarray}
where $|A \rangle$ denotes the initial state of the incoming nucleus A, and $|X \rangle$ represent the final hadronic (or partonic) states.
The sum $\sum_X$ runs over all possible final states except the produced hard quark jet and the radiated gluon.
$J_\mu = \bar{\psi}_i \gamma_\mu  (1-\gamma^5) V_{ij} \psi_j $ is the charged current and $V_{i j}$ is the Cabibbo-Kobayashi-Maskawa flavor mixing matrix~\cite{Aivazis:1993kh}.
In the following, we concentrate on the hadronic tensor $W^{\mu\nu}$ which contains the detailed information about the final state interaction between the propagating quark jet and the traversed nuclear medium.

\begin{figure}[thb]
\includegraphics[width=0.99\linewidth]{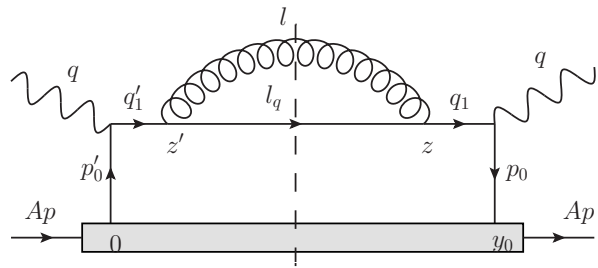}
 \caption{Leading twist contribution to gluon emission from a heavy (or light) quark.
} \label{gluon0}
\end{figure}

Figure~\ref{gluon0} shows the leading twist DIS process in which a $W$ boson carrying momentum $q$ strikes a light (massless) quark from a nucleus constituent at location $y_0'=0$ ($y_0$ in the complex conjugate) and produces a heavy (massive) quark with mass $M$. Here the light quark from the nucleus carries momentum $p_0'$ ($p_0$ in the complex conjugate) and the struck heavy quark carries momentum $q_1'$ ($q_1$ in the complex conjugate).
The heavy quark then emits a real gluon with momentum $l$ and exits the medium without further interaction.
The final outgoing heavy quark carries momentum $l_q$. For the DIS process in Figure~\ref{gluon0}, the hadronic tensor can be written as follows:
\begin{widetext}
\begin{eqnarray}
\label{W0GH}
{W_{(1)}^{A \mu\nu} } &=&  \frac{1}{N_c} \int \frac{d^4l}{(2\pi)^4} (2\pi)\delta(l^2) \int \frac{d^4l_q}{(2\pi)^4} (2\pi) \delta(l_q^2-M^2)
\int d^4y_0 e^{iq\cdot y_0}   \int d^4z \int d^4z'\int \frac{d^4 q_1}{(2\pi)^4} \int \frac{d^4 q_1'}{(2\pi)^4}
\nonumber\\
&\times&
e^{-i q_1 \cdot (y_0-z)}e^{-i q_1' \cdot (z'-y_0')} e^{-i l_q \cdot (z-z')} e^{-i l\cdot (z-z')}
 |V_{ij}|^2 \langle A |\bar{\psi}_i(y_0)  \gamma^\mu (1-\gamma^5) \frac{- i(\slashed{q}_1 + M)}{q_1^2 - M^2 - i\epsilon}
\nonumber\\
&\times&
(- i g \gamma_\alpha T^{a}) (\slashed{l}_q + M)(i g \gamma_\beta T^{a'}) \frac{i(\slashed{q}_1' + M)}{q_1'^2 - M^2 + i\epsilon}  (1+\gamma^5)\gamma^\nu  \psi_i(0)|A\rangle
[\delta_{aa'} \tilde{G}^{\alpha\beta} (l)],
\end{eqnarray}
where $\tilde{G}^{\alpha\beta} (l)$ is the sum of the gluon polarizations.
It is noted that we use full QCD Feynman rules, thus all physical gluon helicity states as well as quark spin states have been properly taken into account.
In the light-cone gauge, $n \cdot A = A^- = 0 $, where the unit four vector is $ n=[1, 0, \mathbf{0}_\perp]$, the polarization sum $\tilde{G}^{\alpha\beta} (l)$ reads:
\begin{eqnarray}
\tilde{G}^{\alpha \beta}(l)&=&-g^{\alpha \beta}+\frac{n^{\alpha}l^{\beta}+n^{\beta}l^{\alpha}}{n\cdot l}.
\end{eqnarray}

In the limit of very high energy, we can ignore the transverse $(\perp)$ component of the quark field operators and factor out one-nucleon state from the nucleus state as follows:
\begin{eqnarray}
\label{factor out aa}
& &\langle A | \bar{\psi}_i(y_0) \gamma^\mu (1-\gamma^5) \hat{O}   (1+\gamma^5)\gamma^\nu \psi_i(0) | A \rangle
\nonumber\\
&\approx&
A C_p^A \langle p | \bar{\psi}_i(y_0^-) \frac{\gamma^+}{2} \psi_i(0) | p \rangle
\frac{1}{4 p^+ q^-} {\rm Tr} [\slashed{p} \gamma^\mu (1-\gamma^5) \{\slashed{q}+ (x_B + x_M)\slashed{p}\} (1+\gamma^5)\gamma^\nu  ]
{\rm Tr} [\frac{\gamma^-}{2}  \hat{O} ],
\end{eqnarray}
where $C_p^A$ denotes the probability of finding a nucleon state with momentum $p$ inside a nucleus (with $A$ nucleons).
Here the momentum fraction $x_M = \frac{M^2}{2 p^+ q^-}$ is defined for convenience.
Now we perform the integration over gluon emission positions $z, z'$ and obtain two $\delta$ functions.  Then we integrate over momenta $q_1, q_1'$, and obtain $q_1 = q_1' = l + l_q $. Also for convenience, we re-introduce the momentum variable $p_0 = l_q + l - q$ and perform the integration over $l_q$, which implies $l_q = q+ p_0 - l = q + p_0' - l$.
After the above simplifications, the hadronic tensor can be organized as follows:
\begin{eqnarray}
{W_{(1)}^{A \mu\nu} } &=&   g^2 C_F \int \frac{d^4l}{(2\pi)^4} (2\pi)\delta(l^2)
\int d^4y_0 \int \frac{d^4 p_0}{(2\pi)^4}e^{- ip_0\cdot y_0}
\frac{1}{q_1^2 - M^2 - i\epsilon}\frac{1}{q_1'^2 - M^2 + i\epsilon}
 (2\pi) \delta(l_q^2-M^2)
 \nonumber\\
&\times&
|V_{ij}|^2 A C_p^A \langle p |\bar{\psi}_i(y_0^-) \frac{\gamma^+}{2} \psi_i(y_0)|p \rangle
\frac{1}{4 p^+ q^-}{\rm Tr} [\slashed{p} \gamma^\mu (1-\gamma^5)  \{\slashed{q}+ (x_B + x_M)\slashed{p}\} (1+\gamma^5) \gamma^\nu ]
\nonumber\\
&\times&{\rm Tr}[\frac{\gamma^-}{2}(\slashed{q}_1 + M)\gamma_\alpha  (\slashed{l}_q + M)
\gamma_\beta (\slashed{q}_1' + M)]\tilde{G}^{\alpha\beta} (l).
\end{eqnarray}

Now we investigate the on-shell condition for the final outgoing heavy quark:
\begin{eqnarray}
\delta(l_q^2-M^2)=\frac{1}{2 p^+ q^-(1-y) }\delta(x_0 - x_B - x_M -\tilde{x}_L ),
\end{eqnarray}
where $y={l^-}/{q^-}$ is the fraction of the forward momentum carried by the radiated gluon with respect to the parent heavy quark. Here for convenience, we also define the momentum fractions $x_0 = p_0^+/p^+$ and $\tilde{x}_L = \frac{l_\perp^2 + y^2 M^2}{2p^+q^-y(1-y)}$.
The above $\delta$ function may be utilized to carry out the integration over $p_0^+ = x_0\, p^+$.
One may further perform the integration over $y_0^+$ and  $\mathbf{y}_{0\,\perp}$, yielding three-dimensional $\delta$ functions for carrying out three-dimensional integrations over $p_0^-$ and $\mathbf{p}_{0\,\perp}$.
Now the differential hadronic tensor for the leading twist DIS process can be expressed as follows:
\begin{eqnarray}
\frac{d W_{(1)}^{A\mu\nu}}{ d y\,d^2\mathbf{l}_\perp} &=&  |V_{ij}|^2 A C_p^A (2\pi) f_i(x_B + x_M + \tilde{x}_L)
\frac{1}{4 p^+ q^-} {\rm Tr} [\slashed{p} \gamma^\mu (1-\gamma^5) \{\slashed{q}+ (x_B + x_M)\slashed{p}\} (1+\gamma^5) \gamma^\nu]
\nonumber\\
&\times&
C_F\frac{\alpha_s}{2 \pi^2} {P(y)}  \frac{l_\perp^2 + \frac{y^4}{1+(1-y)^2}{M^2}}{(l_\perp^2  + y^2 {M^2})^2}.
\end{eqnarray}
In the above expression, $f_i(x)$ is the light quark distribution function in the nucleon of the incoming nucleus,
\begin{eqnarray}
f_i(x) = \int \frac{dy_0^-}{2\pi} e^{-ixp^+y_0^-} \langle p| \bar{\psi}_i(y_0^-) \frac{\gamma^+}{2} \psi_i(0)|p \rangle.
\end{eqnarray}
where $x$ is the fraction of the forward momentum carried by the light quark from the nucleon, and $P(y) = [1+(1-y)^2]/{y}$ is the leading order light quark to gluon (photon) splitting function (Note that the color factor $C_F$ for light quark to gluon splitting vertex have been factor out).
From the above expression, one may read off the differential single gluon emission spectrum from a heavy quark jet in vacuum as:
\begin{eqnarray}
\frac{d N_g^{\rm vac}}{d y\,d^2 \mathbf{l}_\perp  } = C_F  \frac{\alpha_s}{2 \pi^2} {P(y)} \frac{l_\perp^2 + \frac{y^4}{1+(1-y)^2}{M^2}}{(l_\perp^2  + y^2 {M^2})^2}.
\end{eqnarray}
One can see that when the mass $M$ of the quark jet is set to be zero, the above formula reduces to the single gluon emission spectrum from a light quark jet in vacuum \cite{Altarelli:1977zs}.
\end{widetext}

\section{Medium-induced gluon emission in dense nuclear matter}

In the previous section, we have studied the DIS process in which a heavy quark is first produced, then emits a gluon and exits the medium without further interaction.
In this section, we consider the medium modification to the above gluon emisison process from a heavy quark jet.
In particular, we study how the rescattering of the heavy quark or the radiated gluon with the medium constituents influences the single gluon emission spectrum.
In this work, we focus on the single rescattering process.

\begin{figure}[thb]
\includegraphics[width=0.99\linewidth]{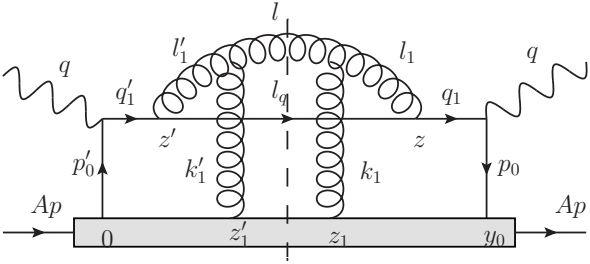}
 \caption{One central-cut diagram for gluon emission from a heavy quark which contributes to the twist-4 hadronic tensor: one rescattering on the radiated gluon in both the amplitude and the complex conjugate.}
\label{gluon1}
\end{figure}

Figure~\ref{gluon1} shows one central-cut diagram describing the process in which the radiated gluon from a heavy quark jet experiences a single rescattering with the medium constituents in both the amplitude and the complex conjugate.
It contributes to the hadronic tensor at the so-called twist-four level. The other 20 diagrams at twist-four level are listed in Appendix A.
In this section, we present the details for the computation of the DIS hadronic tensor and the medium-induced single gluon emission spectrum for Figure~\ref{gluon1}.
The computations for the other 20 diagrams are completely analogous and we list their key results in Appendix A.

In Figure~\ref{gluon1}, a $W$ boson carrying momentum $q$ strikes a light quark from the nucleus at the location $y_0'=0$ ($y_0$ in the complex conjugate) and produces a heavy quark with mass $M$.
The light quark from the nucleus carries momentum $p_0'$ ($p_0$ in the complex conjugate) and the struck heavy quark carries momentum $q_1'$ ($q_1$ in the complex conjugate).
The produced heavy quark propagates through the dense nuclear medium and emits a gluon with momentum $l'_1$ ($l_1$ in the complex conjugate) at the location $z'$ ($z$ in the complex conjugate).
The radiated gluon then scatters off the gluon field of the medium constituent at the location $z_1'$ ($z_1$ in the complex conjugate) and exchanges momentum $k_1'$ ($k_1$ in the complex conjugate) with the nuclear medium.
The final emitted real gluon carries momentum $l$, and the final outgoing heavy quark carries momentum $l_q$.
The hadronic tensor for the DIS process in Figure~\ref{gluon1} can be expressed as follows:
\begin{widetext}
\begin{eqnarray}
 W^{A\mu\nu}_{(\ref{gluon1})}&=& \frac{1}{N_c} \int \frac{d^4 l}{{(2 \pi)}^4}2\pi \delta (l^2)\int \frac{d^4l_q}{{(2\pi)}^4}2\pi \delta (l_q^2-M^2)
 \int d^4 y_0 e^{i q \cdot y_0}\int d^4 z   \int \frac{d^4 q_1}{{(2\pi)}^4}e^{-i q_1 \cdot (y_0-z)}
\nonumber
\\ &\times&
 \int d^4 z_1 \int \frac{d^4 l_1}{{(2\pi)}^4}e^{-i l_1 \cdot (z-z_1)} \int d^4 z_1'\int \frac{d^4 l_1'}{{(2\pi)}^4}e^{-i l_1' \cdot (z_1'-z')} \int d^4 z'\int \frac{d^4 q_1'}{{(2\pi)}^4}e^{-i q_1' \cdot (z'-y_0')} e^{-i l \cdot (z_1-z_1')} e^{-i l_q \cdot (z-z')}
 \nonumber\\ &\times&
|V_{ij}|^2  \langle A | \bar{\psi}_i(y_0)  \gamma^\mu (1-\gamma^5)\frac{-i (\slashed{q}_1+M)}{q_1^2-M^2-i \epsilon}(-i g \gamma_{\alpha_0} T^{a_0}) (\slashed{l}_q+M) \frac{- i \delta_{a_0 b_1}\tilde{G}^{\alpha_0 \beta_1}(l_1)}{l_1^2-i \epsilon}
 \nonumber\\ &\times&
	\left[g f^{b_1 c_1 a_1}\Gamma_{\beta_1 \gamma_1 \alpha_1}(-l_1, -k_1,l) A_{c_1}^{\gamma_1}(z_1)\right]
 \left[\delta _{a_1 a'_1}\tilde{G}^{\alpha_1 \alpha'_1}(l)\right] \left[g f^{a'_1 c'_1 b'_1}\Gamma_{\alpha'_1 \gamma'_1 \beta'_1}(-l, k'_1,l'_1)  A_{c'_1}^{\gamma'_1}(z'_1)\right]
 \nonumber\\ &\times&
	\frac{i \delta_{b'_1 a'_0 } \tilde{G}^{\beta'_1 \alpha'_0 }(l'_1)}{l_1'^2+i \epsilon}(i g\gamma_{\alpha'_0}T^{a'_0}) \frac{i(\slashed{q}'_1+M)}{q_1'^2-M^2+i \epsilon} (1+\gamma^5)\gamma^\nu  \psi_i(0)|A \rangle. 
\label{eq1}
\end{eqnarray}
Here $f^{b_1 c_1 a_1}$ is the anti-symmetric structure constant of the $SU(3)$ color group, and ${\Gamma}_{\beta_1 \gamma_1 \alpha_1}$ is the kernel of three-gluon vetex,
\begin{eqnarray}
{\Gamma}_{\beta_1 \gamma_1 \alpha_1}(-l_1, -k_1,l)
= g_{\beta_1 \gamma_1}{(-l_1+k_1)}_{\alpha_1} + g_{\gamma_1 \alpha_1}{(-k_1-l)}_{\beta_1} + g_{ \alpha_1 \beta_1}{(l+l_1)}_{\gamma_1} \,.
\end{eqnarray}

Now we simplify the hadronic tensor $W^{A\mu\nu}_{(\ref{gluon1})}$. First, one may isolate the phase factors associated with two gluon insertions, $e^{-i(l_q + l_1 - q_1) \cdot z}  e^{i(l_q + l'_1 - q'_1) \cdot z'}$.
Integrating out the coordinate variables $z$ and $z'$, one may obtain two $\delta$ functions, which can be utilized to carry out the integrations over the momenta $q_1$ and $q'_1$ and obtain two momentum-conservation relations: $q_1= l_q + l_1$ and $q'_1= l_q + l'_1$ at vertices $z$ and $z'$.
From momentum conservations in Figure~\ref{gluon1}, we may also obtain the following relations for various momenta,
\begin{eqnarray}
p_0=q_1-q,\,\,\,\, k_1=l-l_1,\,\,\,\, p'_0=q'_1-q,\,\,\,\, k'_1=l-l'_1.
\end{eqnarray}
Here for convenience, we re-introduce the momentum variable $p_0$, with $p_0= l_q + l - q - k_1$ from momentum conservation.
We also change the integration variables $l_1 \to k_1$ and $l'_1 \to k'_1$.
Then the phase factor for $W^{A\mu\nu}_{(\ref{gluon1})}$ can be expressed as: $e^{-ip_0\cdot y_0} e^{-ik_1\cdot z_1} e^{ik'_1\cdot z'_1} $.

In the limit of very high energy, one may factor out the quark field and the gluon field operators from the nucleus state and ignore the transverse $(\perp)$ component of the quark field operators,
\begin{eqnarray}
\label{factor out aa2}
& &\langle A | \bar{\psi}_i(y_0) \gamma^\mu (1-\gamma^5)  \hat{O}  (1+\gamma^5) \gamma^\nu \psi_i(0) | A \rangle
\\&\approx&
A C_p^A \langle p | \bar{\psi}_i(y_0^-) \frac{\gamma^+}{2} \psi_i(0) | p \rangle
\frac{1}{4 p^+ q^-} {\rm Tr} [\slashed{p} \gamma^\mu (1-\gamma^5)  \{\slashed{q}+ (x_B + x_M)\slashed{p}\} (1+\gamma^5) \gamma^\nu ]
{\rm Tr} [\frac{\gamma^-}{2}  \langle A |\hat{O} | A \rangle ].\nonumber
\end{eqnarray}
Then the gluon propagators together with the three-gluon vertices can be simplified as follows:
\begin{eqnarray}
&&\tilde{G}^{\alpha_0 \beta_1}(l_1) {\Gamma}_{\beta_1 \gamma_1 \alpha_1}(-l_1, -k_1,l) A_{c_1}^{\gamma_1}(z_1) \tilde{G}^{\alpha_1 \alpha'_1}(l){\Gamma}_{\alpha_1' \gamma_1' \beta'_1}(-l, k'_1,l'_1)A_{c'_1}^{\gamma'_1}(z'_1)\tilde{G}^{\beta'_1 \alpha'_0 }(l'_1)
\nonumber \\
&& = \tilde{G}^{\alpha_0 \beta_1}(l_1) \left[g_{\alpha_1 \beta_1} (l+l_1)^- A_{c_1}^+(z_1)\right]
\tilde{G}^{\alpha_1 \alpha'_1}(l) \left[g_{\alpha_1' \beta_1'} (l+l_1')^- A_{c_1'}^+(z_1') \right] \tilde{G}^{\beta'_1 \alpha'_0 }(l'_1).
\end{eqnarray}
Here we have only kept the dominant forward $(+)$ component of the scattered gluon field in very high energy limit.
With the above simplifications, the hadronic tensor for Figure~\ref{gluon1} may be written as:
\begin{eqnarray}
 W^{A\mu\nu}_{(\ref{gluon1})}&=& g^4 \int \frac{d^4 l}{{(2 \pi)}^4}2\pi \delta (l^2)\int \frac{d^4l_q}{{(2\pi)}^4}(2\pi)^4 \delta^4(l+l_q - p_0 - k_1 - q)
 \int d^4 y_0  \int d^4 z_1 \int d^4 z'_1
\nonumber\\ 	&\times&
\int \frac{d^4p_0}{(2\pi)^4}\int \frac{d^4 k_1}{{(2\pi)}^4} \int \frac{d^4 k_1'}{{(2\pi)}^4} \left(e^{-ip_0\cdot y_0} e^{-ik_1\cdot z_1} e^{ik_1'\cdot z_1'}\right) \frac{1}{q^2_1-M^2-i\epsilon}\frac{(l+l_1)^-}{l^2_1-i\epsilon}
\frac{1}{q'^2_1-M^2+i\epsilon}
\nonumber\\ &\times&
\frac{(l +l_1')^-}{l'^2_1+i\epsilon}(2\pi) \delta (l_q^2 - M^2)\times |V_{ij}|^2 A C_p^A \langle p | \bar{\psi}_i(y_0^-) \frac{\gamma^+}{2} \psi_i(0) | p \rangle\times\langle A | A_{c_1}^+ (z_1)  A_{c'_1}^+ (z'_1)| A \rangle
 \nonumber\\ 	&\times&
  \frac{1}{4 p^+ q^-} {\rm Tr} [\slashed{p} \gamma^\mu (1-\gamma^5) \{\slashed{q}+ (x_B + x_M)\slashed{p}\}  (1+\gamma^5) \gamma^\nu ]\times \frac{1}{N_c}{\rm Tr}[T^{a_0}f^{a_0 c_1 a_1}f^{a_1 c'_1 a'_0} T^{a'_0}]	
 \nonumber\\ 	&\times&
{\rm Tr}\left[\frac{\gamma^-}{2} (\slashed{q}_1+ M) \gamma_{\alpha_0} (\slashed{l}_q + M) \gamma_{\alpha_0'}(\slashed{q}_1'+ M)\right]
\tilde{G}^{\alpha_0 \beta_1}(l_1) g_{\alpha_1 \beta_1} \tilde{G}^{\alpha_1 \alpha'_1}(l) g_{\alpha_1' \beta_1'} \tilde{G}^{\beta'_1 \alpha'_0 }(l'_1).
\label{eq3}
\end{eqnarray}

We now look at the internal quark and gluon propagators and the final external quark line,
\begin{eqnarray}
& &q_1^2 - M^2 = (q + p_0)^2 -M^2 =  2 p^+ q^- (1 + x_0^-)[-{x}_B + x_0 - \zeta_{M0}],
\\
& &l_1^2 = (l-k_1)^2 = 2p^+q^-(y - \lambda_1^-)[\tilde{x}_L (1-y) - y\, x_M- \lambda_1 - \lambda_{D1}],
\\
& &l_q^2-M^2= (q_1+ k_1 -l)^2 - M^2=2p^+q^-(1+x_0^- + \lambda_1^- -y )[-{x}_B + x_0 +\lambda_1 - \tilde{x}_L (1-y) + y\, x_M - \eta_{M1}],
\end{eqnarray}
where for convenience we have defined the momentum fractions,
\begin{eqnarray}
&& x_0 = \frac{p_0^+}{p^+}, \;\;\;\lambda_1= \frac{k_1^+}{p^+}, \;\;\;
x_0^- =\frac{p_0^-}{q^-}, \;\;\; \lambda_1^- = \frac{k_1^-}{q^-}, \;\;\;
\nonumber\\
&& \zeta_{M0}  = \frac{ {p}_{0\perp}^2+M^2}{2p^+q^-(1 + x_0^-)},\;\;\;  \lambda_{D1} = \frac{(\mathbf{l}_{\perp}-\mathbf{k}_{1 \perp})^2}{2p^+q^-(y-\lambda_1^-)},\;\;\;\eta_{M1} = \frac{(\mathbf{l}_{\perp}-\mathbf{k}_{1 \perp}- \mathbf{p}_{0 \perp})^2+M^2}{2p^+q^-(1+x_0^- + \lambda_1^- -y )}.
\end{eqnarray}
Similarly, we may also define the momentum fractions $x_0'$, $\lambda_1'$, $x_0'^-$, $\lambda_1'^-$, $\zeta_{M0}'$, $\lambda_{D1}'$ and $\eta_{M1}'$.
From the above relations, the contributions from the internal quark and gluon propagator and the on-shell condition for the final outgoing heavy quark can be obtained as:
\begin{eqnarray}
D_q &=&\frac{C_q }{8(p^+)^5(q^-)^3} \frac{y- \frac{\lambda_1^-}{2}}{y-\lambda_1^-}\frac{y- \frac{\lambda_1'^-}{2}}{y-\lambda_1'^-}
\frac{1}{-{x}_B+x_0-\zeta_{M0} -i\epsilon}  \frac{1}{\tilde{x}_L (1-y) - y\, x_M - \lambda_1 - \lambda_{D1}-i\epsilon}
\\&\times&
 \frac{1}{-{x}_B+x_0' -\zeta_{M0}'+i\epsilon}\frac{1}{\tilde{x}_L (1-y) - y\, x_M - \lambda_1' - \lambda_{D1}'+i\epsilon}
 (2\pi)\delta[-{x}_B + x_0 +\lambda_1 - \tilde{x}_L (1-y) + y\, x_M - \eta_{M1}],\nonumber
\end{eqnarray}
where
\begin{eqnarray}
C_q=\frac{1}{(1+x_0^-)(1+x_0'^-)(1+x_0^- +\lambda_1^- - y)}.
\end{eqnarray}
Performing the trace part of the hadronic tensor [the last line of Eq.~(\ref{eq3})], we obtain:
\begin{eqnarray}
	N_q &=& \frac{4 q^-}{C_q}\frac{1+\left(1-\frac{y- \lambda_1^-}{1+x_0^-}\right)\left(1-\frac{y- \lambda_1'^-}{1+x_0'^-}\right)}{(y - \lambda_1^-)(y -\lambda_1'^-)\left(1 -\frac{y- \lambda_1^-}{1+x_0^-}\right)\left(1-\frac{y- \lambda_1'^-}{1+x_0'^-}\right)}
\nonumber\\ &\times&
	 \left[\left(\mathbf{l}_\perp-\mathbf{k}_{1 \perp}-\frac{y- \lambda_1^-}{1+x_0^-}\mathbf{p}_{0 \perp}\right)\cdot\left(\mathbf{l}_\perp-\mathbf{k}_{1 \perp}'-\frac{y- \lambda_1'^-}{1+x_0'^-}\mathbf{p}_{0 \perp}'\right)+\frac{\left(\frac{y - \lambda_1^-}{1+x_0^-}\right)^2\left(\frac{y -\lambda_1'^-}{1+x_0'^-}\right)^2 M^2}{1+\left(1-\frac{y- \lambda_1^-}{1+x_0^-}\right)\left(1-\frac{y- \lambda_1'^-}{1+x_0'^-}\right)}\right].
\end{eqnarray}
With the above simplifications, the hadronic tensor for Figure~\ref{gluon1} now reads:
\begin{eqnarray}
 W^{A\mu\nu}_{(\ref{gluon1})}&=& \frac{g^4}{16 \pi^3}
	\int\frac{dy}{y}  \int d^2 \mathbf{l}_\perp
  \int d y_0^-  \int d z_1^- \int d z_1'^- \int d^3 y_0 \int d^3 z_1 \int d^3 z_1'
\nonumber\\ &\times&
\int \frac{d x_0 }{2\pi}\int \frac{ d \lambda_1}{{2\pi}} \int \frac{ d \lambda_1'}{{2\pi}}\frac{e^{-ix_0 p^+ y_0^-}}{-{x}_B+x_0-\zeta_{M0} -i\epsilon} \frac{e^{-i\lambda_1 p^+  z_1^-}}{\tilde{x}_L (1-y) - y\, x_M - \lambda_1 - \lambda_{D1}-i\epsilon}
\nonumber\\ &\times&
\frac{e^{i\lambda_1' p^+z_1'^-}}{\tilde{x}_L (1-y) - y\, x_M- \lambda_1' - \lambda_{D1}'+i\epsilon}\frac{1}{-{x}_B+x_0' -\zeta_{M0}'+i\epsilon}(2\pi)\delta[-{x}_B + x_0 +\lambda_1 - \tilde{x}_L (1-y) +y\, x_M - \eta_{M1}]
\nonumber\\ &\times&
\int \frac{d^3 \mathbf{p}_0 }{{(2\pi)}^3} \int \frac{d^3 \mathbf{k}_1 }{{(2\pi)}^3} \int \frac{d^3 \mathbf{k}_1' }{{(2\pi)}^3}
 e^{-i\mathbf{p}_0\cdot \mathbf{y}_0}e^{-i\mathbf{k}_1\cdot \mathbf{z}_1} e^{i\mathbf{k}_1'\cdot \mathbf{z}_1'}
|V_{ij}|^2 A C_p^A \langle p | \bar{\psi}_i(y_0^-) \frac{\gamma^+}{2} \psi_i(0) | p \rangle
\nonumber\\&\times&
\frac{1}{4 p^+ q^-} {\rm Tr} [\slashed{p} \gamma^\mu (1-\gamma^5) \{\slashed{q}+ (x_B + x_M)\slashed{p}\} (1+\gamma^5)\gamma^\nu  ]
\langle A | A_{c_1}^+ (z_1)  A_{c'_1}^+ (z'_1)| A \rangle  \frac{1}{N_c}{\rm Tr}[T^{a_0}f^{a_0 c_1 a_1}f^{a_1 c'_1 a'_0} T^{a'_0}]
\nonumber\\ &\times&
\frac{1}{2(p^+q^-)^2}\frac{y- \frac{\lambda_1^-}{2}}{y-\lambda_1^-}\frac{y- \frac{\lambda_1'^-}{2}}{y-\lambda_1'^-}\frac{1+\left(1-\frac{y- \lambda_1^-}{1+x_0^-}\right)\left(1-\frac{y- \lambda_1'^-}{1+x_0'^-}\right)}{(y - \lambda_1^-)(y -\lambda_1'^-)\left(1 -\frac{y- \lambda_1^-}{1+x_0^-}\right)\left(1-\frac{y- \lambda_1'^-}{1+x_0'^-}\right)}
\nonumber\\ &\times&
	 \left[\left(\mathbf{l}_\perp-\mathbf{k}_{1 \perp}-\frac{y- \lambda_1^-}{1+x_0^-}\mathbf{p}_{0 \perp}\right)\cdot\left(\mathbf{l}_\perp-\mathbf{k}_{1 \perp}'-\frac{y- \lambda_1'^-}{1+x_0'^-}\mathbf{p}_{0 \perp}'\right)+\frac{\left(\frac{y - \lambda_1^-}{1+x_0^-}\right)^2\left(\frac{y -\lambda_1'^-}{1+x_0'^-}\right)^2 M^2}{1+\left(1-\frac{y- \lambda_1^-}{1+x_0^-}\right)\left(1-\frac{y- \lambda_1'^-}{1+x_0'^-}\right)}\right].
\end{eqnarray}
Here for convenience, the three-vector notations for coordinate and momentum are used: $\mathbf{z} = (z^+, \mathbf{z}_{\perp})$ and $\mathbf{k} = (k^-, \mathbf{k}_{\perp})$. Their dot product reads: $\mathbf{k} \cdot \mathbf{z} = k^-z^+ - \mathbf{k}_{\perp} \cdot \mathbf{z}_{\perp}$.

Now we perform the integrations over the momentum fractions $x_0$, $\lambda_1$ and $\lambda_1'$.
The $\delta$ function from the on-shell condition of the outgoing heavy quark can be used to carry out the integration over $x_0$,
\begin{eqnarray}
&&\int \frac{dx_0}{2 \pi}\frac{e^{- i x_0 p^+ y_0^-}}{-{x}_B+x_0-\zeta_{M0} -i\epsilon}\frac{1}{-{x}_B+x_0'-\zeta_{M0}' +i\epsilon}(2\pi)\delta[-{x}_B + x_0 +\lambda_1 - \tilde{x}_L (1-y) + y\, x_M - \eta_{M1}]
\\
	&&= e^{- i ({x}_B +\tilde{x}_L (1-y) - y\, x_M + \eta_{M1} ) p^+ y_0^-} \frac{e^{+ i \lambda_1 p^+ y_0^-}}{\tilde{x}_L (1-y) - y\, x_M - \lambda_1 + \eta_{M1} - \zeta_{M0}-i\epsilon}\frac{1}{\tilde{x}_L (1-y) - y\, x_M - \lambda_1' + \eta_{M1}' - \zeta_{M0}'+i\epsilon}.\,\,\nonumber
\end{eqnarray}
The integration over $\lambda_1$ may be preformed by closing the contour with a counterclockwise semicircle in the upper half of the $\lambda_1$ complex plane:
\begin{eqnarray}
	&&\int \frac{d\lambda_1}{2 \pi}\frac{e^{- i \lambda_1 p^+ (z_1^--y_0^-)}}{[\tilde{x}_L (1-y) - y\, x_M - \lambda_1 + \eta_{M1} - \zeta_{M0}-i\epsilon][\tilde{x}_L (1-y) - y\, x_M - \lambda_1 - \lambda_{D1}-i\epsilon]}
\nonumber\\&&\,\, =\,\, i \theta(z_1^- -y_0^-)e^{-i (\tilde{x}_L (1-y) - y\, x_M) p^+ (z_1^- - y_0^-)}e^{i (\eta_{M1} - \zeta_{M0}) p^+ y_0^-} e^{i \lambda_{D1} p^+ z_1^-} \frac{e^{- i \chi_{M10} p^+ y_0^-} - e^{- i \chi_{M10} p^+ z_1^-} }{\chi_{M10}},
\end{eqnarray}
where the momentum fraction $\chi_{M10} = \eta_{M1} +\lambda_{D1} - \zeta_{M0}$ has been defined for convenience.
The integration over the momentum fraction $\lambda_1'$ (in the complex conjugate) is completely analogous.
After carrying out the integration over the momentum fractions $x_0$, $\lambda_1$ and $\lambda_1'$, the hadronic tensor for Figure~\ref{gluon1} reads:
\begin{eqnarray}
 W^{A\mu\nu}_{(\ref{gluon1})}&=& \frac{g^4}{16 \pi^3}\int\frac{dy}{y}  \int d^2 \mathbf{l}_\perp
	\frac{1}{4 p^+ q^-} {\rm Tr} [\slashed{p} \gamma^\mu (1-\gamma^5) \{\slashed{q}+ (x_B + x_M)\slashed{p}\} (1+\gamma^5)\gamma^\nu  ]
\nonumber\\ &\times&
\int d y_0^- \int d^3 y_0 \int \frac{d^3 \mathbf{p}_0 }{{(2\pi)}^3} e^{-i\mathbf{p}_0\cdot \mathbf{y}_0} e^{- i ({x}_B +x_M + \tilde{x}_L ) p^+ y_0^-}
|V_{ij}|^2 A C_p^A
	\langle p | \bar{\psi}_i(y_0^-) \frac{\gamma^+}{2} \psi_i(0) | p \rangle
\nonumber\\ &\times&
\int d z_1^- \int d z_1'^- [i \theta(z_1^- -y_0^-)][- i\theta(z_1'^- -y_0'^-)]
 \int d^3 z_1 \int d^3 z_1'
	\int \frac{d^3 \mathbf{k}_1 }{{(2\pi)}^3} \int \frac{d^3 \mathbf{k}_1' }{{(2\pi)}^3} e^{-i\mathbf{k}_1\cdot \mathbf{z}_1} e^{i\mathbf{k}_1'\cdot \mathbf{z}_1'}
 \nonumber\\ &\times&
	\langle A | A_{c_1}^+ (z_1)  A_{c'_1}^+ (z'_1)| A \rangle  \frac{1}{N_c}{\rm Tr}[T^{a_0}f^{a_0 c_1 a_1}f^{a_1 c'_1 a'_0} T^{a'_0}]
\nonumber\\ &\times&
e^{i (\tilde{x}_L + x_M -\zeta_{M0}) p^+ y_0^-} e^{- i (\tilde{x}_L (1-y) - y\, x_M-\lambda_{D1}) p^+ z_1^-}
e^{i (\tilde{x}_L (1-y) - y\, x_M-\lambda_{D1}') p^+ z_1'^-}
 \nonumber\\ &\times&
[e^{- i \chi_{M10} p^+ y_0^-} - e^{- i \chi_{M10} p^+ z_1^-}][1 - e^{i \chi_{M10}' p^+ z_1'^-}]
 \nonumber\\ &\times&
\frac{1}{\chi_{M10}}\frac{1}{\chi'_{M10}}\frac{1}{2(p^+q^-)^2}\frac{y- \frac{\lambda_1^-}{2}}{y-\lambda_1^-}\frac{y- \frac{\lambda_1'^-}{2}}{y-\lambda_1'^-}\frac{1+\left(1-\frac{y- \lambda_1^-}{1+x_0^-}\right)\left(1-\frac{y- \lambda_1'^-}{1+x_0'^-}\right)}{(y - \lambda_1^-)(y -\lambda_1'^-)\left(1 -\frac{y- \lambda_1^-}{1+x_0^-}\right)\left(1-\frac{y- \lambda_1'^-}{1+x_0'^-}\right)}
\nonumber\\ &\times&
	 \left[\left(\mathbf{l}_\perp-\mathbf{k}_{1 \perp}-\frac{y- \lambda_1^-}{1+x_0^-}\mathbf{p}_{0 \perp}\right)\cdot\left(\mathbf{l}_\perp-\mathbf{k}_{1 \perp}'-\frac{y- \lambda_1'^-}{1+x_0'^-}\mathbf{p}_{0 \perp}'\right)+\frac{\left(\frac{y - \lambda_1^-}{1+x_0^-}\right)^2\left(\frac{y -\lambda_1'^-}{1+x_0'^-}\right)^2 M^2}{1+\left(1-\frac{y- \lambda_1^-}{1+x_0^-}\right)\left(1-\frac{y- \lambda_1'^-}{1+x_0'^-}\right)}\right].
\end{eqnarray}

Now we look at the correlations of two gluon field operators in the nucleus state $  \langle A |  A_{c_1}^+(z_1) A_{c_1'}^+(z_1')  |A\rangle$. One may first factor out the color factor for the gluon field correlator as follows:
\begin{eqnarray}
 \langle A |  A_{c_1}^+(z_1) A_{c_1'}^+(z_1')  |A\rangle
	= \frac{1}{d(R)} {\rm Tr} [T_{c_1}(R) T_{c_1'}(R)] \langle A |  A^+(z_1) A^+(z_1')  |A\rangle
	= \delta_{c_1 c_1'} \frac{C_2(R)}{N_c^2-1} \langle A| A^+(z_1) A^+(z_1') | A\rangle,
\end{eqnarray}
where $d(R)$ and $C_2(R)$ are the dimensions and the Casimir factors for the representation $R$ of the $SU(3)$ color group.
If the exchanged gluon field is initiated by the on-shell quark, we have $d(R)=3$ and $C_2(R) = C_F$.
If the exchanged gluon field is initiated by the on-shell gluon, $d(R)=8$ and $C_2(R) = C_A$.
Since the two gluon insertions from the amplitude and the complex conjugate carry the same color ($\delta_{c_1 c_1'}$), one may evaluate the color factor for the hadronic tensor $W^{A\mu\nu}_{(\ref{gluon1})}$ as follows:
\begin{eqnarray}
	\delta_{c_1 c_1'}\frac{1}{N_c} {\rm Tr}[T^{a_0}f^{a_0 c_1 a_1}f^{a_1 c_1' a_0'} T^{a_0'}] = C_F C_A.
\end{eqnarray}
In addition, we change the coordinate variables $(z_1, z_1')$ to their mean value and difference $(Z_1, \delta z_1)$:
 \begin{eqnarray}
Z_1 = \frac{z_1 + z_1'}{2},\;\;\;\;\;\delta z_1 = z_1 - z_1'.
\end{eqnarray}
By using the translational invariance, the gluon field correlations only depend on $\delta z_1$,
\begin{eqnarray}
\label{translational invariance}
\langle A| A^+(z_1) A^+(z'_1) | A\rangle \approx \langle A| A^+(\delta{z_1}) A^+(0) | A\rangle.
\end{eqnarray}
Now we may carry out the following integration for the phase factor,
\begin{eqnarray}
\int d^3\mathbf{z}_1 \int d^3 \mathbf{z}'_1 e^{-i\mathbf{k}_1 \cdot \mathbf{z}_1} e^{i\mathbf{k}'_1\cdot \mathbf{z}'_1}
= (2\pi)^3 \delta^3(\mathbf{k}_1 - \mathbf{k}'_1) \int d^3 \mathbf{\delta z}_1 e^{-i(\mathbf{k}_1 + \mathbf{k}'_1) \cdot \frac{\delta\mathbf{z}_1}{2}}.
\end{eqnarray}
The above $\delta$ function comes from the integration over the three-coordinate $\mathbf{Z}_1$ (the mean value of two gluon insertions), and means that two gluon field insertions carry the same momentum, $\mathbf{k}'_1 = \mathbf{k}_1$.
We also perform the integration over the three-coordinate $\mathbf{y}_0$, which gives $\mathbf{p}_0=0$, thus $\mathbf{p}'_0 = \mathbf{p}_0 + \mathbf{k}_1 - \mathbf{k}'_1 = 0$.
With the above simplifications, the hadronic tensor for Figure~\ref{gluon1} may be written as:
\begin{eqnarray}
\label{eq50}
	W^{A\mu\nu}_{(\ref{gluon1})} &=& \frac{g^4}{16 \pi^3}
	\int\frac{dy}{y}  \int d^2 \mathbf{l}_\perp  |V_{ij}|^2 A C_p^A (2 \pi)f_i(x_B + x_M + \tilde{x}_L)
\frac{1}{4 p^+ q^-}{\rm Tr} [\slashed{p} \gamma^\mu (1-\gamma^5) \{\slashed{q}+(x_B + x_M)\slashed{p}\}  (1+\gamma^5) \gamma^\nu ]
	\nonumber\\ &\times&
 \int d Z_1^- \int d\delta z_1^-  \int d^3 \mathbf{\delta z}_1\int \frac{d^3\mathbf{k}_1}{(2\pi)^3} e^{-i\mathbf{k}_1\cdot \delta \mathbf{z}_1} \frac{ C_2(R)}{N_c^2-1}  \langle A | A^+(\delta z_1^-, \delta \mathbf{z}_1) A^+(0)  |A\rangle
\nonumber \\ &\times&
e^{i (\tilde{x}_L + x_M -\zeta_{M0}) p^+ y_0^-} e^{- i (\tilde{x}_L (1-y) - y\, x_M-\lambda_{D1}) p^+ \delta z_1^-}(e^{- i \chi_{M10} p^+ y_0^-} - e^{- i \chi_{M10} p^+ (Z_1^- +\frac{1}{2} \delta z_1^-)})(1 - e^{i \chi_{M10} p^+ (Z_1^- -\frac{1}{2} \delta z_1^-)})
 \nonumber\\ &\times&
C_F C_A \frac{1}{{(\chi_{M10})}^2}\frac{2}{{(2 p^+ q^-)}^2}\frac{1+{\left(1 + \lambda_1^- - y\right)}^2}{{(y - \lambda_1^-)}^2{\left(1 + \lambda_1^- -y \right)}^2}
{\left(\frac{y-\frac{\lambda_1^-}{2}}{y- \lambda_1^-}\right)}^2 \left[\left(\mathbf{l}_\perp-\mathbf{k}_{1 \perp}\right)^2+\frac{(y - \lambda_1^-)^4 M^2}{1+(1+ \lambda_1^- - y)^2}\right] .
\end{eqnarray}

Now we simplify the second last line of Eq.~(\ref{eq50}), which we refer to as the phase factor $S_{(\ref{gluon1})}$.
We first note that the coordinate $Z_1^-$ is the mean value of two gluon insertion points and can span over the whole nucleus size, while the coordinates $y_0^-$ and $\delta z_1^-$ are confined within the nucleon size.
Therefore, we may suppress the contributions from $y_0^-$ and $ \delta z_1^- $ as compared to the contribution from $Z_1^-$.
Then the phase factor $S_{(\ref{gluon1})}$ can be greatly simplified,
\begin{eqnarray}
\label{deltas2}
S_{(\ref{gluon1})} &=&  e^{i (\tilde{x}_L + x_M -  \zeta_{M0}) p^+ y_0^-} e^{- i (\tilde{x}_L (1-y) - y\, x_M-\lambda_{D1}) p^+ \delta z_1^-}(e^{- i \chi_{M10} p^+ y_0^-} - e^{- i \chi_{M10} p^+ (Z_1^- +\frac{1}{2} \delta z_1^-)})(1 - e^{i \chi_{M10} p^+ (Z_1^- -\frac{1}{2} \delta z_1^-)})
\nonumber\\  &\approx&
2-2 \cos(\chi_{M10} p^+ Z_1^-).
\end{eqnarray}
Recalling the definitions of the momentum fractions $\eta_{M1}$, $\lambda_{D1}$ and $\zeta_{M0}$, one may obtain the momentum fraction variable $\chi_{M10} = \eta_{M1} + \lambda_{D1} - \zeta_{M0}$ as follows:
\begin{eqnarray}
\chi_{M10}=
&=&\frac{{\left(\mathbf{l}_\perp - \mathbf{k}_{1 \perp} \right)}^2+(y-\lambda_1^-)^2 M^2}{2p^+ q^-(y - \lambda_1^-)\left(1+\lambda_1^- - y\right)}
= \tilde{x}_L \;\frac{y (1-y)}{(y - \lambda_1^-)(1+\lambda_1^--y)} \frac{{\left(\mathbf{l}_\perp - \mathbf{k}_{1 \perp} \right)}^2+(y-\lambda_1^-)^2 M^2}{l_\perp^2 + y^2 M^2}.
\end{eqnarray}
The above expression may be used to evaluate the last line of Eq.~(\ref{eq50}), which we refer to as the hard matrix element $T_{(2)}$,
\begin{eqnarray}
\label{deltat2}
	T_{(\ref{gluon1})} &=& {2 y P(y)} C_F C_A \left[\frac{1+(1+ \lambda_1^- -y)^2}{1+(1-y)^2} {\left(\frac{y-\frac{\lambda_1^-}{2}}{y- \lambda_1^-}\right)}^2 \frac{{\left(\mathbf{l}_\perp - \mathbf{k}_{1 \perp} \right)}^2+\frac{(y- \lambda_1^-)^4 M^2}{1+(1+ \lambda_1^- -y)^2}}{\left[{\left(\mathbf{l}_\perp - \mathbf{k}_{1 \perp} \right)}^2+(y-\lambda_1^-)^2 M^2\right]^2}\right] = {2 y P(y)} \tilde{T}_{(\ref{gluon1})},
\end{eqnarray}
where we have also defined the kernel of the hard matrix element $\tilde{T}_{(\ref{gluon1})}$ for convenience.
With the above simplifications, the differential hadronic tensor for Figure~\ref{gluon1} now reads:
\begin{eqnarray}
\frac{d W^{A\mu\nu}_{(\ref{gluon1})}}{d y d^2 \mathbf{l}_\perp} &=& |V_{ij}|^2 A C_p^A (2 \pi)f_i(x_B + x_M + \tilde{x}_L)
\frac{1}{4 p^+ q^-} {\rm Tr} [\slashed{p} \gamma^\mu (1-\gamma^5)  \{\slashed{q}+ (x_B + x_M)\slashed{p}\} (1+\gamma^5) \gamma^\nu ]
\nonumber\\ &\times&
\frac{\alpha_s}{2 \pi^2}P(y)\int d Z_1^-
\int \frac{d^3\mathbf{k}_1}{(2\pi)^3}\int d\delta z_1^- \int d^3 \mathbf{\delta z}_1  e^{-i\mathbf{k}_1\cdot \delta \mathbf{z}_1}
\left(g^2 \frac{C_F C_2(R)}{N_c^2-1}\right)\langle A | A^+(\delta z_1^-, \delta \mathbf{z}_1) A^+(0)  |A\rangle
\nonumber\\ &\times&
\left[2-2 \cos\left(\frac{y (1-y)}{(y - \lambda_1^-)(1+\lambda_1^--y)} \frac{{\left(\mathbf{l}_\perp - \mathbf{k}_{1 \perp} \right)}^2+(y-\lambda_1^-)^2 M^2}{l_\perp^2 + y^2 M^2}\,\frac{Z_1^-}{{\tilde{\tau}_{\rm form}}^-}\right)\right]
\nonumber\\ &\times&
C_A \left[\frac{1+(1+ \lambda_1^- -y)^2}{1+(1-y)^2} {\left(\frac{y-\frac{\lambda_1^-}{2}}{y- \lambda_1^-}\right)}^2 \frac{{\left(\mathbf{l}_\perp - \mathbf{k}_{1 \perp} \right)}^2+\frac{(y- \lambda_1^-)^4 M^2}{1+(1+ \lambda_1^- -y)^2}}{\left[{\left(\mathbf{l}_\perp - \mathbf{k}_{1 \perp} \right)}^2+(y-\lambda_1^-)^2 M^2\right]^2}\right],
\end{eqnarray}
where ${\tilde{\tau}_{\rm form}}^- = {1}/{(\tilde{x}_L p^+)}$ is the formation time for gluon radiation from a heavy quark jet.
We note that the above equation can reduce to Eq. (39) in our previous work \cite{Zhang:2018kkn} when we take zero mass for the hard quark jet.
Now we define the following distribution function $\mathcal{D}(k_1^-,\mathbf{k}_{1 \perp})$ for the exchanged three-momentum $\mathbf{k}_1=(k_1^-,\mathbf{k}_{1 \perp})$ with the nuclear medium,
\begin{eqnarray}
\mathcal{D}(k_1^-,\mathbf{k}_{1 \perp}) = \int d \delta z_1^- \int d^3 \delta \mathbf{z}_{1} e^{-i \mathbf{k}_{1} \cdot \delta \mathbf{z}_{1}}\left( g^2 \frac{C_F C_2(R)}{N_c^2-1}\right)\langle A | A^+(\delta z_1^-, \delta \mathbf{ z}_{1}) A^+(0)  |A\rangle.\;\;\;\;\;\;\;\;\;
\end{eqnarray}
One can show that in the high energy limit, the distribution function $\mathcal{D}(k_1^-,\, \mathbf{k}_{1\, \perp})$ is, up to a constant factor, the differential elastic scattering rate ${dP_{\rm el}}/{dk_1^- d^2\mathbf{k}_{1\perp} dZ_1^-}$ between a light quark and medium constituents (see Appendix B),
\begin{eqnarray}
\mathcal{D}(k_1^-,\, \mathbf{k}_{1\, \perp}) 
= (2\pi)^3 \frac{dP_{\rm el}}{dk_1^- d^2\mathbf{k}_{1\perp} dZ_1^-}.
\end{eqnarray}
One may read off the medium-induced single gluon emission spectrum for Figure~\ref{gluon1} as follows:
\begin{eqnarray}
\frac{d N^{med}_{g,\,(\ref{gluon1})}}{d y d^2 \mathbf{l}_\perp} &=&
\frac{\alpha_s}{2 \pi^2}P(y)\int d Z_1^-  \int \frac{d^3\mathbf{k}_1}{(2\pi)^3}\mathcal{D}(k_1^-,\, \mathbf{k}_{1\, \perp})
\nonumber\\ &\times&
\left[2-2 \cos\left(\frac{y (1-y)}{(y - \lambda_1^-)(1+\lambda_1^--y)} \frac{{\left(\mathbf{l}_\perp - \mathbf{k}_{1 \perp} \right)}^2+(y-\lambda_1^-)^2 M^2}{l_\perp^2 + y^2 M^2}\,\frac{Z_1^-}{{\tilde{\tau}_{\rm form}}^-}\right)\right]
\nonumber\\ &\times&
C_A \left[\frac{1+(1+ \lambda_1^- -y)^2}{1+(1-y)^2} {\left(\frac{y-\frac{\lambda_1^-}{2}}{y- \lambda_1^-}\right)}^2 \frac{{\left(\mathbf{l}_\perp - \mathbf{k}_{1 \perp} \right)}^2+\frac{(y- \lambda_1^-)^4 M^2}{1+(1+ \lambda_1^- -y)^2}}{\left[{\left(\mathbf{l}_\perp - \mathbf{k}_{1 \perp} \right)}^2+(y-\lambda_1^-)^2 M^2\right]^2}\right].
\end{eqnarray}
The calculations for the other $20$ diagrams are completely analogous and we list their main results together with the diagrams in Appendix A.
After summing up the contributions from all $21$ diagrams, we obtain the following expression for the medium-induced single gluon emission spectrum from a heavy quark jet interacting with the nuclear medium via single rescattering:
\begin{eqnarray}
\frac{d N^{med}_{g}}{d y d^2 \mathbf{l}_\perp} &=&
\frac{\alpha_s}{2 \pi^2}P(y)\int d Z_1^-  \int \frac{d k_1^- d^2\mathbf{k}_{1\perp}}{(2\pi)^3} \mathcal{D}(k_1^-,\, \mathbf{k}_{1\, \perp})
\nonumber\\ &\times&
\left\{\left[2-2 \cos\left( \frac{y(1-y)}{(y - \lambda_1^-)\left(1+\lambda_1^--y\right)}\frac{{\left(\mathbf{l}_{\perp}- \mathbf{k}_{1 \perp} \right)}^2 + (y - \lambda_1^-)^2 M^2}{l_{\perp}^2 + y^2 M^2} \frac{Z_1^-}{{\tilde{\tau}_{\rm form}}^-} \right)\right]\right.
\nonumber\\ && \left.
\times
C_A \left[\frac{1+{(1 + \lambda_1^- - y) }^2}{1+{\left(1  - y \right)}^2}
{\left(\frac{y-\frac{\lambda_1^-}{2}}{y- \lambda_1^-}\right)}^2
\nonumber\right.
\frac{{\left(\mathbf{l}_{\perp} - \mathbf{k}_{1 \perp} \right)}^2 + \frac{(y - \lambda_1^-)^4  M^2}{1+(1 + \lambda_1^- -y)^2} }{\left[{\left(\mathbf{l}_{\perp} - \mathbf{k}_{1 \perp} \right)}^2 + (y - \lambda_1^-)^2 M^2\right]^2}
\right.\nonumber \\ & &
-\frac{1+{(1 + \lambda_1^- - y )}(1 - y )}{2[1+{\left(1  - y \right)}^2]}
{\frac{y-\frac{\lambda_1^-}{2}}{y- \lambda_1^-}}
\frac{{\mathbf{l}_{\perp} \cdot \left(\mathbf{l}_{\perp} - \mathbf{k}_{1 \perp} \right)} + \frac{y^2(y - \lambda_1^-)^2  }{1+(1 + \lambda_1^- -y)(1 -y)} M^2}{\left[{l_{\perp}^2} + y^2 {M^2}\right]\left[{{\left(\mathbf{l}_{\perp} - \mathbf{k}_{1 \perp} \right)}^2 + (y - \lambda_1^-)^2 M^2}\right]}
 \nonumber\\ & &\left.
- \frac{1+{(1 + \lambda_1^- - y )}{\left(1  - \frac{y}{1 + \lambda_1^-} \right)}}{2[1+{\left(1  - y \right)}^2]}
{\frac{y-\frac{\lambda_1^-}{2}}{y- \lambda_1^-}}
	\frac{\left(\mathbf{l}_{\perp} -\mathbf{k}_{1 \perp} \right)\cdot \left(\mathbf{l}_{\perp} - \frac{y}{1 + \lambda_1^-}\mathbf{k}_{1 \perp} \right) + \frac{\left(\frac{y}{1 + \lambda_1^-}\right)^2(y - \lambda_1^-)^2  }{1+(1 + \lambda_1^- -y)\left(1  - \frac{y}{1 + \lambda_1^-} \right)} M^2}{\left[\!{\left(\mathbf{l}_{\perp} - \frac{y}{1 + \lambda_1^-} \mathbf{k}_{1 \perp} \right)}^2\! + \!\left(\frac{y}{1 + \lambda_1^-} \right)^2 \! M^2 \right] \! \left[{\left(\mathbf{l}_{\perp} - \mathbf{k}_{1 \perp} \right)}^2\! + (y - \lambda_1^-)^2 M^2\right]}
\right]\nonumber\\ & &
+ \left[2 - 2\cos \left( \frac{Z_1^-}{{\tilde{\tau}_{\rm form}}^-} \right)\right]
\left[
C_F \frac{{l}_{\perp}^2 +\frac{y^4}{1+(1-y)^2}M^2}{\left[{l}_{\perp}^2 +y^2 M^2\right]^2}
- \frac{C_A}{2}
\frac{\left(y - \frac{\lambda_1^-}{2}\right)^2}{y (y - \lambda_1^-)} \frac{{l}_{\perp}^2 +\frac{y^4}{1+(1-y)^2}M^2}{\left[{l}_{\perp}^2 +y^2 M^2\right]^2}
\nonumber\right.\\& &\left.
+
\left(\frac{C_A}{2}-C_F\right)\frac{1+{(1 - y )}\left(1 - \frac{y}{1 + \lambda_1^-} \right)}{1+{\left(1  - y \right)}^2}\frac{{\mathbf{l}_{\perp} \cdot \left(\mathbf{l}_{\perp} - \frac{y}{1 + \lambda_1^-}\mathbf{k}_{1 \perp} \right)} + \frac{y^2\left(\frac{y}{1 + \lambda_1^-}\right)^2  M^2}{1+{(1 - y )}\left(1 - \frac{y}{1 + \lambda_1^-} \right)} }{\left[{l_{\perp}^2} + y^2 {M^2}\right]\left[{\left(\mathbf{l}_{\perp} - \frac{y}{1 + \lambda_1^-}\mathbf{k}_{1 \perp} \right)}^2 +\left(\frac{y}{1 + \lambda_1^-}\right)^2 M^2\right]}
\right]
 \nonumber\\ & &
 \left.
 +C_F\left[\frac{1+{\left(1 -\frac{ y }{1 + \lambda_1^-}\right)}^2}{1+{\left(1  - y \right)}^2}
\frac{{{\left(\mathbf{l}_{\perp} - \frac{y}{1 + \lambda_1^-}\mathbf{k}_{1 \perp} \right)}^2 + \frac{\left(\frac{y}{1 + \lambda_1^-}\right)^4  }{1+{\left(1 -\frac{ y }{1 + \lambda_1^-}\right)}^2}M^2 }}{\left[{\left(\mathbf{l}_{\perp} -\frac{y}{1 + \lambda_1^-} \mathbf{k}_{1 \perp} \right)}^2 + \left(\frac{y}{1 + \lambda_1^-} \right)^2 M^2\right]^2} -\frac{{l}_{\perp}^2 +\frac{y^4}{1+(1-y)^2}M^2}{\left[{l}_{\perp}^2 +y^2 M^2\right]^2}\right]\right\}.
\end{eqnarray}
The above formula is quite general in the sense that the property of the nuclear medium that the hard jet parton probes is contained in the distribution function $\mathcal{D}(k_1^-,\, \mathbf{k}_{1\, \perp})$.
In other words, as we work beyond the collinear rescattering expansion, the medium-induced emission spectrum is controlled by the full distribution of the differential elastic scattering rate ${dP_{\rm el}}/{dk_1^- d^2\mathbf{k}_{1\perp} dZ_1^-}$ between the light quarks and medium constituents (in the high energy limit).
In addition, the contributions from both transverse and longitudinal momentum exchanges with the medium constituents are included in the above  medium-induced single gluon emission spectrum.
Compared to our previous work \cite{Zhang:2018kkn}, here we express our medium-induced gluon emission spectrum in terms of the distribution function $\mathcal{D}(k_1^-, \mathbf{k}_{1\perp})$ which contains the detailed information about the dense nuclear medium probed by the jet partons.

In the following, we model the traversed nuclear medium by heavy static scattering centers and take the static screened potential for the exchanged gluon field \cite{Sievert:2018imd},
\begin{eqnarray}
\label{yukawa potential2}
	 A^{\mu}(\mathbf{p})=g^{\mu - } (2 \pi)\delta(p^-)\frac{- g}{\mathbf{p}_{\perp}^2+\mu^2},
\end{eqnarray}
where $\mu$ is the screened mass of the exchanged gluon. Note that the color factor of the gluon field has been factored out.
In the above static potential, $A^+$ component is nonzero and $p^-$ component is set to be zero. Compared to our previous work \cite{Zhang:2018kkn}, this is a more consistent setup as we have used the light-cone coordinate throughout our calculation. Also in the high energy limit, we only keep the dominant (+) component of the scattered gluon field.
For hard partons scattering with the above static potential, only transverse momenta are exchanged between the hard partons and medium constituents.
To include the contribution from longitudinal momentum (and/or energy) exchange between jet and medium, one may consider the dense nuclear medium consisting of dynamical constituents \cite{Arnold:2001ba, Arnold:2002ja, Djordjevic:2007at, Djordjevic:2008iz}.
By performing the Fourier transformation, the gluon field correlation may be obtained:
\begin{eqnarray}
\label{AA IN CARTESIAN2}
\langle A | A^\mu(\delta \mathbf{z}_1) A^\nu(0)| A \rangle &=&
\delta_+^\mu \delta_+^\nu \rho^- \delta(\delta z_1^-)\int \frac{d^2 \mathbf{p}_\perp }{(2 \pi)^2}e^{i \mathbf{p}_\perp \cdot {\delta \mathbf{z}_{1 \perp}}} \frac{g^2}{(\mathbf{p}_\perp^2 + \mu^2)^2},\;\;\;\;\;\;
\end{eqnarray}
where $\rho^-$ is the light-cone density of the medium constituents (scattering centers) that the hard jet parton interacts.
One may also compute the differential elastic cross section for a light quark scattering with the above static screened potential as follows:
\begin{eqnarray}
\frac{d\sigma_{\rm el}}{d^2 \mathbf{p}_\perp} = \frac{ C_F  C_2(R) }{N_c^2-1}\frac{ |g A^+(\mathbf{p}_{\perp}) |^2}{4\pi^2} = \frac{ C_F C_2(R)}{N_c^2-1} \frac{4\alpha_s^2}{(\mathbf{p}_\perp^2 + \mu^2)^2}.\;\;\;\;\;\;\;
\end{eqnarray}
Using the above expressions, the distribution function $\mathcal{D}(k_1^-,\, \mathbf{k}_{1\, \perp})$ can be obtained as follows:
\begin{eqnarray}
\mathcal{D}(k_1^-,\,\mathbf{k}_{1\, \perp})&=& (2\pi)\delta(k_1^-) (2 \pi)^2  \rho^-\frac{d\sigma_{\rm el}}{d^2 \mathbf{k}_{1\,\perp}} = (2\pi)\delta(k_1^-) \mathcal{D}_{\perp}(\mathbf{k}_{1\, \perp}).
\end{eqnarray}
Here for convenience, we have also defined the distribution function $\mathcal{D}_{\perp}(\mathbf{k}_{1\, \perp})$ for transverse momentum exchange,
\begin{eqnarray}
\mathcal{D}_{\perp}(\mathbf{k}_{1\, \perp}) & = & (2 \pi)^2  \rho^-\frac{d\sigma_{\rm el}}{d^2 \mathbf{k}_{1\,\perp}}
= (2 \pi)^2 \frac{d P_{\rm el}}{d^2 \mathbf{k}_{1\,\perp}dZ_1^-},
\end{eqnarray}
where ${dP_{\rm el}}/{(d^2 \mathbf{k}_{1\,\perp}dZ_1^-)}$ is the differential elastic scattering rate with only transverse momentum exchange.
Given the above distribution function, the light quark (light-cone) transport coefficient $\hat{q}_{lc}$ may be computed,
\begin{eqnarray}
	\hat{q}_{lc} = \frac{d\langle k_{1\perp}^2\rangle}{dL^-} = \int \frac{dk_1^- d^2\mathbf{k}_{1\perp}}{(2\pi)^3} \mathbf{k}_{1\, \perp}^2 \mathcal{D}(k_1^-,\,\mathbf{k}_{1\, \perp}) =\int \frac{d^2\mathbf{k}_{1\perp}}{(2\pi)^2} \mathbf{k}_{1\, \perp}^2 \mathcal{D}_\perp(\mathbf{k}_{1\, \perp}) =\int {d^2\mathbf{k}_{1\perp}} \mathbf{k}_{1\, \perp}^2 \rho^-\frac{d\sigma_{\rm el}}{d^2 \mathbf{k}_{1\,\perp}}.
\end{eqnarray}
Using the above static screened potential for the exchanged gluon field, the medium-induced single gluon emission spectrum can be simplified as follows:
\begin{eqnarray}
\frac{d {N_g^{\rm med}}}{d y d^2 \mathbf{l}_\perp}&=& \frac{\alpha_s}{2\pi^2} {P(y)}  \int {dZ_1^-}
\int \frac{d^2\mathbf{k}_{1\perp}}{(2\pi)^2} \mathcal{D}_{\perp}(\mathbf{k}_{1\, \perp})
\nonumber
\\ &\times&
\left\{C_A\left[2-2 \cos\left(\frac{{\left(\mathbf{l}_{\perp}- \mathbf{k}_{1 \perp} \right)}^2 + y^2 M^2}{l_{\perp}^2 + y^2 M^2} \frac{Z_1^-}{{\tilde{\tau}_{\rm form}}^-} \right)\right] \times\left[\frac{{\left(\mathbf{l}_{\perp} - \mathbf{k}_{1 \perp} \right)}^2 + \frac{y^4 }{1+(1 -y)^2} M^2 }{\left[{\left(\mathbf{l}_{\perp} - \mathbf{k}_{1 \perp} \right)}^2 + y^2 M^2\right]^2}
\nonumber\right.\right.\\ & &\left.
- \frac{1}{2}
\frac{{\mathbf{l}_{\perp} \cdot \left(\mathbf{l}_{\perp} - \mathbf{k}_{1 \perp} \right)} + \frac{y^4}{1+(1-y)^2} M^2}{\left[{l_{\perp}^2} + y^2 {M^2}\right]\left[{{\left(\mathbf{l}_{\perp} - \mathbf{k}_{1 \perp} \right)}^2 + y^2 M^2}\right]}
-\frac{1}{2}\frac{\left(\mathbf{l}_{\perp} -\mathbf{k}_{1 \perp} \right)\cdot \left(\mathbf{l}_{\perp} - {y}\mathbf{k}_{1\perp} \right) + \frac{{y}^4}{1+(1 -y)^2} M^2}{\left[{\left(\mathbf{l}_{\perp} - y \mathbf{k}_{1 \perp} \right)}^2 + {y}^2 M^2\right]\left[{\left(\mathbf{l}_{\perp} - \mathbf{k}_{1 \perp} \right)}^2 + y^2 M^2\right]}
\right]
\nonumber\\ & &
+ \left(
\frac{C_A}{2}-C_F\right)\left[2 - 2\cos \left( \frac{Z_1^-}{{\tilde{\tau}_{\rm form}}^-} \right)\right]\left[
\frac{{\mathbf{l}_{\perp} \cdot \left(\mathbf{l}_{\perp} - y\mathbf{k}_{1 \perp} \right)} + \frac{y^4 }{1+(1 -y)^2}  M^2}{\left[{l_{\perp}^2} + y^2 {M^2}\right]\left[{\left(\mathbf{l}_{\perp} - y \mathbf{k}_{1 \perp} \right)}^2 +y^2 M^2\right]}-\frac{{l}_{\perp}^2 +\frac{y^4}{1+(1-y)^2}M^2}{\left[{l}_{\perp}^2 +y^2 M^2\right]^2} \right]
 \nonumber\\& &\left. +C_F\left[
\frac{{{\left(\mathbf{l}_{\perp} - y \mathbf{k}_{1 \perp} \right)}^2 + \frac{y^4 }{1+{(1 -y)}^2}M^2 }}{\left[{\left(\mathbf{l}_{\perp} -y \mathbf{k}_{1 \perp} \right)}^2 + y^2 M^2\right]^2} -\frac{{l}_{\perp}^2 +\frac{y^4}{1+(1-y)^2}M^2}{\left[{l}_{\perp}^2 +y^2 M^2\right]^2}\right]\right\}.
\end{eqnarray}
The above formula represents the medium-induced single gluon emission spectrum from a heavy quark jet when interacting with the static screened potential.
Since only transverse momentum is exchanged with the nuclear medium, the gluon emission spectrum is controlled by the distribution of transverse momentum exchange ${D}_\perp(\mathbf{k}_{1\, \perp})$, or the differential elastic scattering rate ${dP_{\rm el}}/{d^2 \mathbf{k}_{1\,\perp}dZ_1^-}$, between the light quarks and the medium constituents.
Note that the above formula is still beyond the collinear rescattering expansion and the soft-gluon emission limit.
It can reduce to Eq. (53) in our previous work~\cite{Zhang:2018kkn} when we take zero mass for the hard quark jet.
One may further perform the simplification by taking the soft-gluon emission limit ($y = l^-/q^- \ll 1$).
For our case, if we take the limit $y^2 M \ll y M \sim l_\perp \sim k_{1\perp}$, the above result can be reduced to the following form:
\begin{eqnarray}
	\frac{d {N_g^{\rm med}}}{d y d^2\mathbf{l}_\perp} &=&
	\frac{\alpha_s}{2\pi^2} {P(y)} \int {dZ_1^-} \int {d^2\mathbf{k}_{1\perp}}
\frac{dP_{\rm el}}{d^2 \mathbf{k}_{1\,\perp}dZ_1^-}
\times C_A\left[2-2 \cos\left(\frac{{\left(\mathbf{l}_{\perp}- \mathbf{k}_{1 \perp} \right)}^2 + y^2 M^2}{l_{\perp}^2 + y^2 M^2} \frac{Z_1^-}{{\tilde{\tau}_{\rm form}}^-} \right)\right]
\\ &\times&
\left[\frac{{\left(\mathbf{l}_{\perp} - \mathbf{k}_{1 \perp} \right)}^2  }{\left[{\left(\mathbf{l}_{\perp} - \mathbf{k}_{1 \perp} \right)}^2 + y^2 M^2\right]^2}-
\frac{{\mathbf{l}_{\perp} \cdot \left(\mathbf{l}_{\perp} - \mathbf{k}_{1 \perp} \right)} }{\left[{l_{\perp}^2} + y^2 {M^2}\right]\left[{{\left(\mathbf{l}_{\perp} - \mathbf{k}_{1 \perp} \right)}^2 + y^2 M^2}\right]}\right]
.\nonumber
\end{eqnarray}
Note that in the small $y$ limit, $P(y)$ reduces to $2/y$ and $\tau_{\rm form}^-$ reduces to $2q^-y/l_\perp^2 = 2l^-/l_\perp^2$.
The above result means that when we only consider the contribution from the transverse scattering and take the limit of soft-gluon emission, our result for medium-induced gluon emission reduces to the first order in opacity DGLV formula (with zero thermal mass for radiated gluon) \cite{Djordjevic:2003zk}.
The above formula will reduce to Eq. (54) in our previous work~\cite{Zhang:2018kkn} (and the GLV result \cite{Gyulassy:1999zd, Gyulassy:2000er}) when zero mass is taken for the hard quark jet.

\end{widetext}

\section{Summary}

In this work, we have studied the medium-induced gluon emission process experienced by a hard quark jet propagating through the dense nuclear matter.
Using the framework of deep inelastic scattering off a large nucleus, we have derived a general formula for the medium-induced single gluon emission from a heavy (or light) quark jet interacting with the dense nuclear medium via both transverse and longitudinal scatterings.
As we have not performed the collinear rescattering expansion, our medium-induced gluon emission spectrum is controlled by the full distribution of the differential elastic scattering rates between the propagating partons and the medium constituents.
Thus, our current work can be viewed as a generalization of the HT formalism for both heavy and light flavor medium-induced radiative process.
We have further shown that our medium-induced gluon radiation result can reduce to the first order in opacity DGLV (with zero effective mass for radiated gluon) formula if one utilizes heavy static scattering centers for the traversed nuclear matter and also takes the soft-gluon emission limit.
The established connection between HT and DGLV formalisms, and more importantly, the above mentioned improvements of our formalism contribute to a significant progress on our understanding of the medium-induced radiative process experienced by the hard quark jet  traversing and interacting with the dense nuclear matter.
The application of our formalism to phenomenological jet quenching studies in the context of high-energy nuclear collisions will be performed in the future effort.

\section*{ACKNOWLEDGMENTS}

We thank X.-N. Wang for discussions.
This work is supported in part by Natural Science Foundation of China (NSFC) under grant Nos. 11775095, 11890711 and 11375072. D.-F.H. is supported by Ministry of Science and Technology of China (MSTC) under ``973" project No. 2015CB856904(4) and by NSFC under grant Nos. 11735007, 11521064, 11375070.

\begin{widetext}
\section*{APPENDIX A}

In this Appendix, we present the main calculation results for the other 20 cut diagrams, as shown in Figures~(\ref{heavy2}-\ref{heavy12}).

\begin{figure}[thb]
\centering
\includegraphics[width=0.99\linewidth]{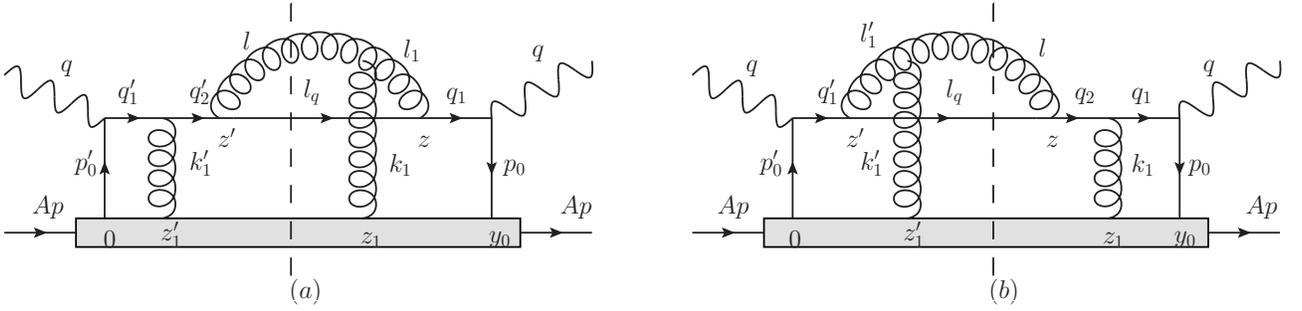}
 \caption{Two central-cut diagrams: one rescattering on the quark before gluon emission in the amplitude and one rescattering on the radiated gluon in the complex conjugate (a) or vice versa (b).}
	\label{heavy2}
\end{figure}
The phase factor for Figure~\ref{heavy2} reads:
\begin{eqnarray}
S_{(\ref{heavy2},a)}&=& e^{- i\zeta_{M0} p^+ y_0^-}  e^{- i (\tilde{x}_L (1-y) - y\, x_M-\lambda_{D1}) p^+ z_1^-} e^{- i \zeta_{M0}' p^+ z_1'^-} e^{i (\tilde{x}_L (1-y) - y\, x_M+\eta_{M1}') p^+ z_1'^-}  \left(e^{- i \chi_{M10} p^+ y_0^-} - e^{- i \chi_{M10} p^+ z_1^-}\right)
\nonumber\\  &\approx&
 e^{i\chi_{M10} p^+ Z_1^-} -1,
\nonumber\\
S_{(\ref{heavy2},b)}&=& e^{ i \zeta_{M0} p^+ z_1^-} e^{-i (\tilde{x}_L (1-y) - y\, x_M+\eta_{M1}) p^+ z_1^-} e^{ i (\tilde{x}_L (1-y) - y\, x_M-\lambda_{D1}') p^+ z_1'^-}   \left(1 - e^{i \chi_{M10}' p^+ z_1'^-}\right)
\nonumber\\  &\approx&
e^{- i\chi_{M10} p^+ Z_1^-} -1,
\nonumber\\
S_{(\ref{heavy2})}&=&S_{(\ref{heavy2},a)}+S_{(\ref{heavy2},b)}\approx
[2\cos(\chi_{M10} p^+ Z_1^-)-2].
\end{eqnarray}
The matrix element for Figure~\ref{heavy2} reads:
\begin{eqnarray}
 \tilde{T}_{(\ref{heavy2},a)} =\tilde{T}_{(\ref{heavy2},b)}
&=& C_F\frac{C_A}{2}\frac{1+\left(1+\lambda_1^- - y \right)\left(1-\frac{y}{1+\lambda_1^-}\right)}{1+(1-y)^2}
\frac{y-\frac{\lambda_1^-}{2}}{y-\lambda_1^-}
\nonumber\\&\times&
\frac{\left(\mathbf{l}_{\perp} -\mathbf{k}_{1 \perp} \right)\cdot \left(\mathbf{l}_{\perp} - \frac{y}{1 + \lambda_1^-}\mathbf{k}_{1 \perp}\right)+\frac{\left(\frac{y}{1+\lambda_1^-}\right)^2(y-\lambda_1^-)^2 }{1+\left(1+\lambda_1^- - y \right)\left(1-\frac{y}{1+\lambda_1^-}\right)}M^2}{\left[\left(\mathbf{l}_{\perp} -\mathbf{k}_{1 \perp} \right)^2 +(y-\lambda_1^-)^2 M^2\right]\left[\left(\mathbf{l}_{\perp} - \frac{y}{1 + \lambda_1^-}\mathbf{k}_{1 \perp}\right)^2+\left(\frac{y}{1 + \lambda_1^-}\right)^2 M^2\right]} .
\end{eqnarray}

\begin{figure}[thb]
\centering
\includegraphics[width=0.99\linewidth]{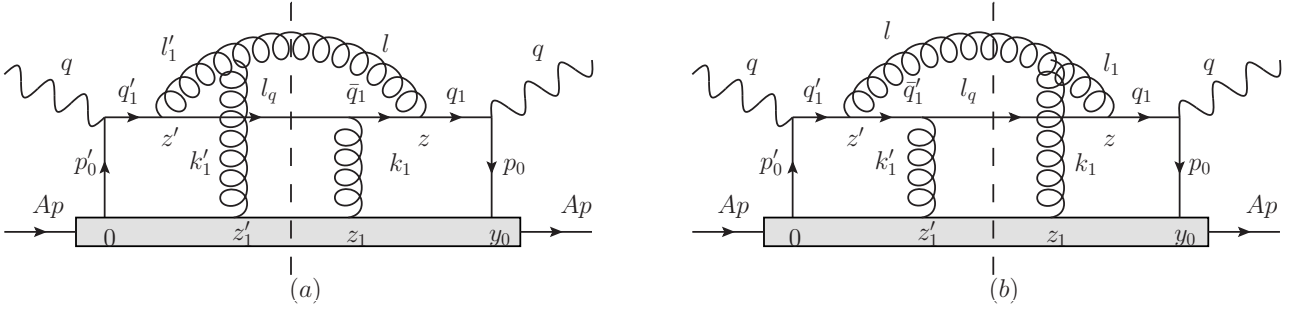}
 \caption{Two central-cut diagrams: one rescattering on the radiated gluon in the amplitude and one rescattering on the quark after gluon emission in the complex conjugate (a) or vice versa (b).}
	\label{heavy3}
\end{figure}
The phase factor for Figure~\ref{heavy3} reads:
\begin{eqnarray}
S_{(\ref{heavy3},a)}&=&  e^{i (\eta_{M0}-\eta_{M1}) p^+ z_1^-}  e^{ i (\tilde{x}_L (1-y) - y\, x_M-\lambda_{D1}') p^+ z_1'^-}  \left(e^{- i \tilde{x}_L p^+ y_0^-} - e^{-i \tilde{x}_L p^+ z_1^-}\right) \left(1 - e^{ i \chi_{M10}' p^+ z_1'^-}\right)
\nonumber\\  &\approx&
1 - e^{i \tilde{x}_L p^+ Z_1^-} - e^{-i \chi_{M10} p^+ Z_1^-} + e^{-i (\chi_{M10}-\tilde{x}_L) p^+ Z_1^-},
\nonumber\\
S_{(\ref{heavy3},b)}&=& e^{-i\zeta_{M0} p^+ y_0^-} e^{- i (\tilde{x}_L (1-y) - y\, x_M-\lambda_{D1}) p^+ z_1^-} e^{- i (\eta_{M0}'-\eta_{M1}') p^+ z_1'^-} \left(e^{- i \chi_{M10} p^+ y_0^-} - e^{- i \chi_{M10} p^+ z_1^-}\right)\left(1 - e^{i \tilde{x}_L p^+ z_1'^-}\right)
\nonumber\\  &\approx&
1 - e^{-i \tilde{x}_L p^+ Z_1^-} - e^{+i \chi_{M10} p^+ Z_1^-} + e^{i (\chi_{M10}-\tilde{x}_L) p^+ Z_1^-},
\nonumber\\
S_{(\ref{heavy3})}&=&S_{(\ref{heavy3},a)}+S_{(\ref{heavy3},b)}\approx
2 -2\cos( \tilde{x}_L p^+ Z_1^-) - 2\cos( \chi_{M10} p^+ Z_1^-) + 2\cos[(\chi_{M10}-\tilde{x}_L) p^+ Z_1^-],
\end{eqnarray}
where
\begin{eqnarray}
\eta_{M0} = \frac{l_{\perp}^2+M^2}{2p^+q^-(1 - y )}.
\end{eqnarray}
The matrix element for Figure~\ref{heavy3} reads:
\begin{eqnarray}
\tilde{T}_{(\ref{heavy3},a)} =\tilde{T}_{(\ref{heavy3},b)}
= - C_F\frac{C_A}{2}\frac{1+\left(1+\lambda_1^- - y \right)\left(1-y\right)}{1+(1-y)^2}
\frac{y-\frac{\lambda_1^-}{2}}{y-\lambda_1^-}
\frac{\mathbf{l}_{\perp} \cdot \left(\mathbf{l}_{\perp}- \mathbf{k}_{1 \perp}\right)+\frac{y^2(y-\lambda_1^-)^2 }{1+\left(1+\lambda_1^- - y \right)(1-y)}M^2}{\left[\left(\mathbf{l}_{\perp}- \mathbf{k}_{1 \perp}\right)^2+(y-\lambda_1^-)^2 M^2\right]\left[l_\perp^2+y^2 M^2\right]}.
\end{eqnarray}

\begin{figure}[thb]
\centering
\includegraphics[width=0.49\linewidth]{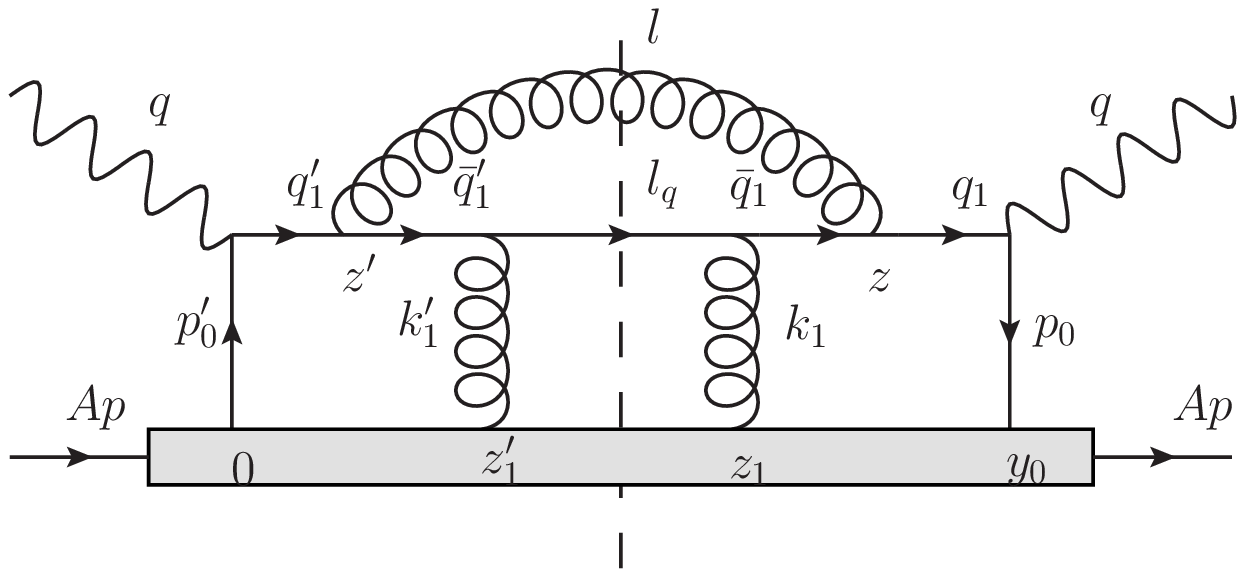}
 \caption{A central-cut diagram: one rescattering on the quark after gluon emission in both the amplitude and the complex conjugate.}
	\label{heavy4}
\end{figure}
The phase factor for Figure~\ref{heavy4} reads:
\begin{eqnarray}
S_{(\ref{heavy4})}&=& e^{-i\zeta_{M0} p^+y_0^-} e^{i (\eta_{M0}-\eta_{M1}) p^+ z_1^-}e^{- i (\eta_{M0}'-\eta_{M1}') p^+ z_1'^-}(e^{-i\tilde{x}_Lp^+y_0^-}-e^{-i\tilde{x}_L p^+ z_1^-})(1 - e^{i\tilde{x}_Lp^+z_1'^-})
\nonumber\\ &\approx&
2-2 \cos(\tilde{x}_L p^+ Z_1^-).
\end{eqnarray}

The matrix element for Figure~\ref{heavy4} reads:
\begin{eqnarray}
\tilde{T}_{(\ref{heavy4})}
&=& C_F^2 \frac{l_\perp^2+\frac{y^4 M^2}{1+(1-y)^2}}{[l_\perp^2 + y^2 M^2]^2}.
\end{eqnarray}

\begin{figure}[thb]
\centering
\includegraphics[width=0.49\linewidth]{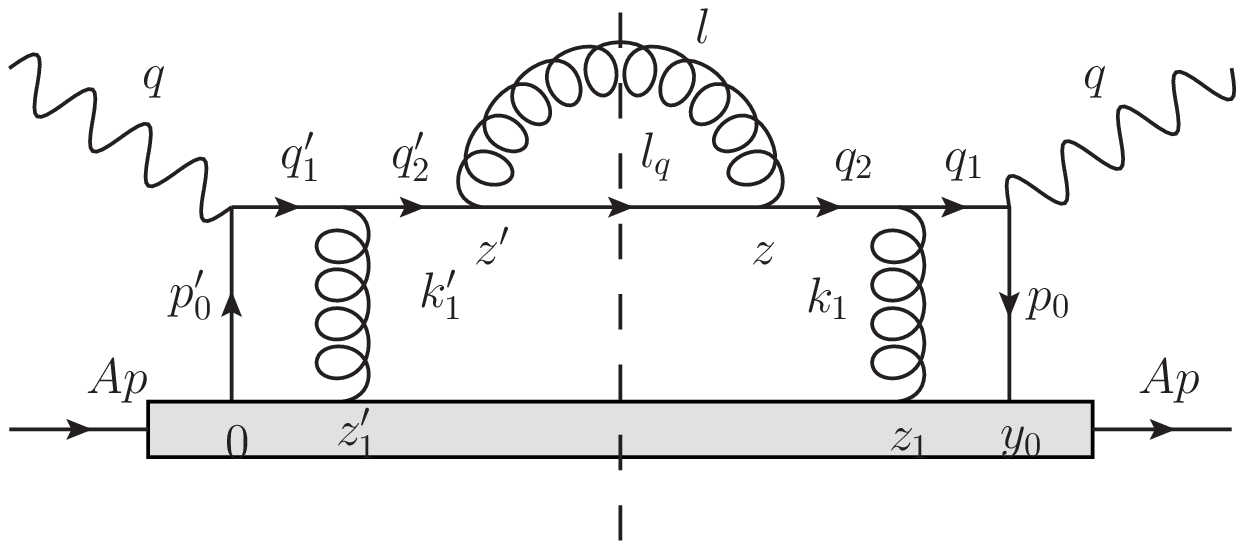}
 \caption{A central-cut diagram: one rescattering on the quark before gluon emission in both the amplitude and the complex conjugate.}
	\label{heavy5}
\end{figure}
The phase factor for Figure~\ref{heavy5} reads:
\begin{eqnarray}
S_{(\ref{heavy5})}&=& e^{-i \zeta_{M0} p^+y_0^-} e^{- i \tilde{x}_L p^+  z_1^-} e^{- i \tilde{x}_L p^+  z_1'^-}
\approx 1.
\end{eqnarray}
The matrix element for Figure~\ref{heavy5} reads:
\begin{eqnarray}
\tilde{T}_{(\ref{heavy5})}
&=& C_F^2\frac{1+\left(1-\frac{y}{1+ \lambda_1^-}\right)^2}{1+(1-y)^2}\frac{\left(\mathbf{l}_{\perp} - \frac{y}{1+ \lambda_1^-} \mathbf{k}_{1\perp}\right)^2+\frac{\left(\frac{y}{1+ \lambda_1^-}\right)^4 M^2}{1+\left(1-\frac{y}{1+ \lambda_1^-}\right)^2}}{\left[\left(\mathbf{l}_{\perp} - \frac{y}{1+ \lambda_1^-} \mathbf{k}_{1\perp}\right)^2 + \left(\frac{y}{1+ \lambda_1^-}\right)^2 M^2\right]^2} .
\end{eqnarray}

\begin{figure}[thb]
\centering
\includegraphics[width=0.99\linewidth]{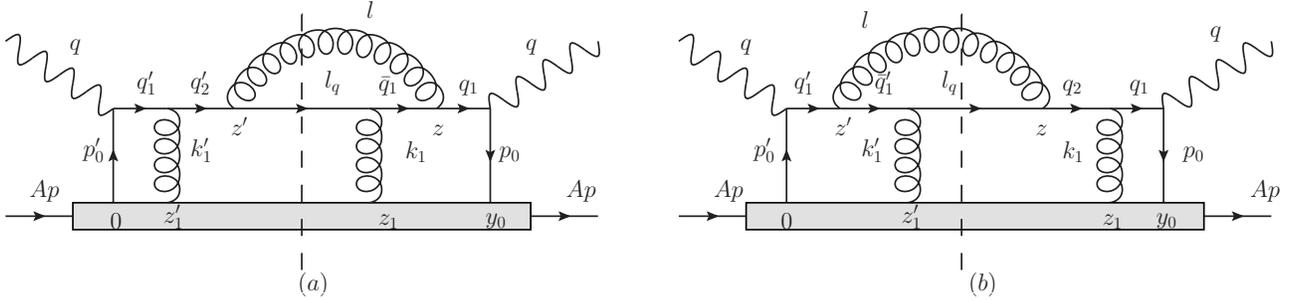}
 \caption{Two central-cut diagrams: one rescattering on the quark before gluon emission in the amplitude and one rescattering on the quark after gluon emission in the complex conjugate (a) or vice versa (b).}
	\label{heavy6}
\end{figure}
The phase factor for Figure~\ref{heavy6} reads:
\begin{eqnarray}
S_{(\ref{heavy6},a)}&=&e^{- i \zeta_{M0} p^+ y_0^-}e^{-i ( \eta_{M1}-\eta_{M0}) p^+ z_1^-}e^{i \tilde{x}_L p^+ z_1'^-}(e^{-i \tilde{x}_L p^+ y_0^-} - e^{-i \tilde{x}_L p^+ z_1^-})
\approx e^{i \tilde{x}_L p^+ Z_1^-}-1,
\nonumber\\
S_{(\ref{heavy6},b)}&=&e^{i ( \eta_{M1}'-\eta_{M0}') p^+ z_1'^-}e^{-i \tilde{x}_L p^+ z_1^-}(1 - e^{i \tilde{x}_L p^+ z_1^-})
\approx e^{-i \tilde{x}_L p^+ Z_1^-}-1,
\nonumber\\
S_{(\ref{heavy6})}&=&S_{(\ref{heavy6},a)}\,\,+\,\,S_{(\ref{heavy6},b)}\,\,\approx\,\,
2 \cos(\tilde{x}_L p^+ Z_1^-) -2.
\end{eqnarray}
The matrix element for Figure~\ref{heavy6} reads:
\begin{eqnarray}
\hspace{-5mm}
\tilde{T}_{(\ref{heavy6},a)}=\tilde{T}_{(\ref{heavy6},b)}
= C_F \left( C_F-\frac{C_A}{2}\right)\frac{1+\left(1 - y \right)\left(1-\frac{y}{1+ \lambda_1^-}\right)}{1+(1-y)^2} \frac{\mathbf{l}_{\perp} \cdot \left(\mathbf{l}_{\perp} - \frac{y}{1+ \lambda_1^-} \mathbf{k}_{1\perp}\right)+\frac{y^2\left(\frac{y}{1+\lambda_1^-}\right)^2 M^2}{1+(1-y)\left(1-\frac{y}{1+\lambda_1^-}\right)}}{
\left[l_\perp^2+y^2 M^2\right]\left[\left(\mathbf{l}_{\perp} - \frac{y}{1+ \lambda_1^-} \mathbf{k}_{1\perp}\right)^2+\left(\frac{y}{1+ \lambda_1^-}\right)^2 M^2\right]}.
\end{eqnarray}

\begin{figure}[thb]
\centering
\includegraphics[width=0.99\linewidth]{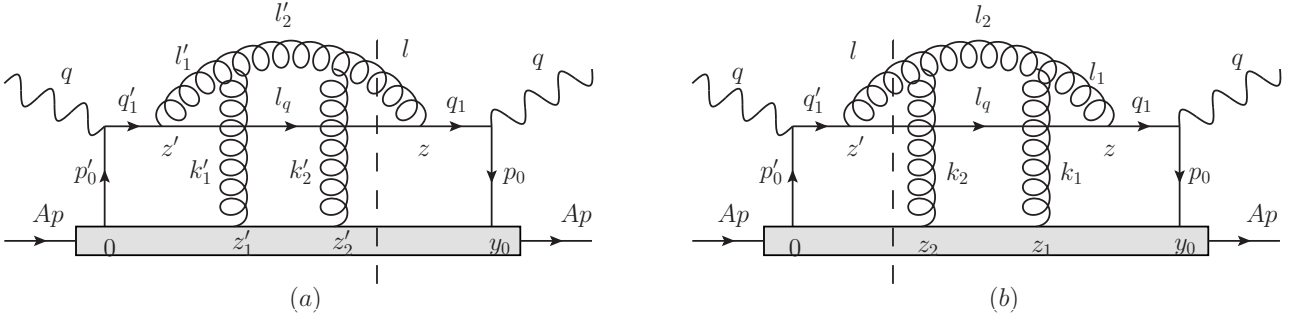}
	\caption{Two non-central-cut diagrams: two rescatterings on the radiated gluons in the amplitude and zero rescattering in the complex conjugate (a) or vice versa (b).}
	\label{heavy7}
\end{figure}
The phase factor for Figure~\ref{heavy7} reads:
\begin{eqnarray}
S_{(\ref{heavy7},a)}&=&\frac{1}{2}e^{- i (\tilde{x}_L (1-y) - y\, x_M-\lambda_{D1}') p^+ z_1'^-}e^{i (\tilde{x}_L (1-y) - y\, x_M-\lambda_{D1}') p^+ z_2'^-}(e^{i \tilde{x}_L p^+ z_1^-}-1)
\approx\frac{1}{2}(e^{i \tilde{x}_L p^+ Z_1^-}-1),
\nonumber\\
S_{(\ref{heavy7},b)}&=&\frac{1}{2}e^{- i \zeta_{M0} p^+ y_0^-}e^{i (\tilde{x}_L (1-y) - y\, x_M-\lambda_{D1}) p^+  z_1^-}e^{-i (\tilde{x}_L (1-y) - y\, x_M-\lambda_{D1}) p^+  z_2^-}(e^{-i \tilde{x}_L p^+ z_1^-}-e^{-i \tilde{x}_L p^+ y_0^-})
\nonumber\\ &\approx&\frac{1}{2}(e^{-i \tilde{x}_L p^+ Z_1^-}-1),
\nonumber\\ S_{(\ref{heavy7})}&=&S_{(\ref{heavy7},a)}\,\,+\,\,S_{(\ref{heavy7},b)}\,\,\approx\,\,
\cos(\tilde{x}_L p^+ Z_1^-) -1.
\end{eqnarray}
The matrix element for Figure~\ref{heavy7} reads:
\begin{eqnarray}
 \tilde{T}_{(\ref{heavy7},a)}=\tilde{T}_{(\ref{heavy7},b)}
 & =& C_F C_A \frac{\left(y -\frac{\lambda_1^-}{2}\right)^2}{y ( y - \lambda_1^-)}\frac{l_\perp^2+\frac{y^4 M^2}{1+(1-y)^2}}{[l_\perp^2 + y^2 M^2]^2}.
\end{eqnarray}

\begin{figure}[thb]
\centering
\includegraphics[width=0.99\linewidth]{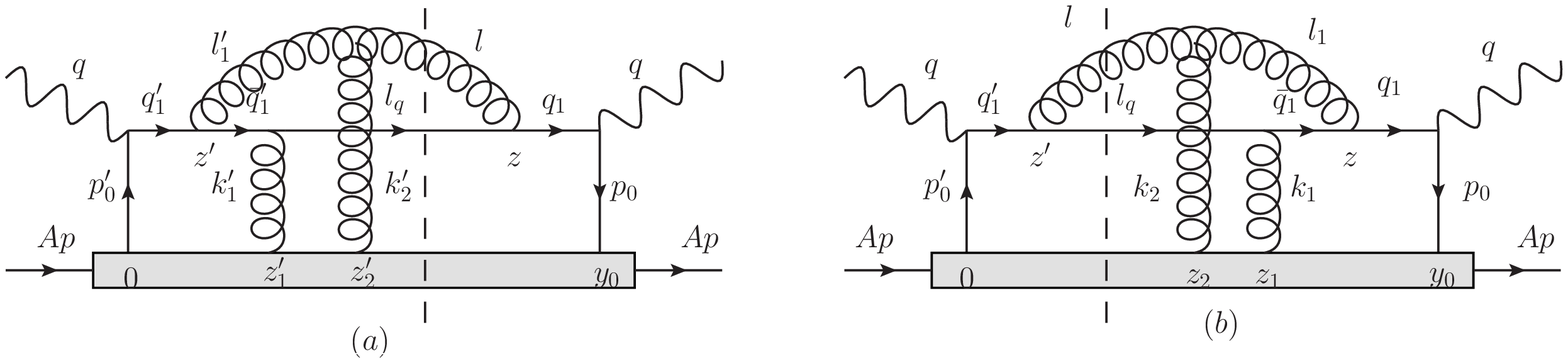}
	\caption{Two non-central-cut diagrams: two rescatterings (with one on the quark after gluon emission and one on the radiated gluon) in the amplitude and zero rescattering in the complex conjugate (a) or vice versa (b).}
	\label{heavy8}
\end{figure}

The phase factor for Figure~\ref{heavy8} reads:
\begin{eqnarray}
S_{(\ref{heavy8},a)}&=&e^{-i (\eta_{M1}'-\eta_{M0}')p^+ z_1'^-}e^{ i (\tilde{x}_L (1-y) - y\, x_M-\lambda_{D1}')p^+ z_2'^-}( -1+\frac{1}{2}e^{i \chi_{M10}' p^+ z_1'^- }+ \frac{1}{2}e^{i \chi_{M10}' p^+ z_2'^- })
\nonumber\\&\approx&
e^{i \tilde{x}_L p^+ Z_1^- } - e^{i (\chi_{M10}-\tilde{x}_L) p^+ Z_1^- },
\nonumber\\
S_{(\ref{heavy8},b)}&=&e^{-i \zeta_{M0}p^+y_0^-}e^{i (\eta_{M1}-\eta_{M0})p^+ z_1^-}e^{- i (\tilde{x}_L (1-y) - y\, x_M-\lambda_{D1})p^+ z_2^-}( -e^{-i \chi_{M10} p^+ y_0^- }+\frac{1}{2}e^{-i \chi_{M10} p^+ z_1^- }+ \frac{1}{2}e^{-i \chi_{M10} p^+ z_2^- })
\nonumber\\&\approx&
e^{-i \tilde{x}_L p^+ Z_1^- } - e^{-i (\chi_{M10}-\tilde{x}_L) p^+ Z_1^- },
\nonumber\\
S_{(\ref{heavy8})}&=&S_{(\ref{heavy8},a)}\,\,+\,\,S_{(\ref{heavy8},b)}\,\,\approx\,\,
2 \cos(\tilde{x}_L p^+ Z_1^-) - 2\cos[(\tilde{x}_L-\chi_{M10}) p^+ Z_1^-].
\end{eqnarray}
The matrix element for Figure~\ref{heavy8} reads:
\begin{eqnarray}
& & \tilde{T}_{(\ref{heavy8},a)}=\tilde{T}_{(\ref{heavy8},b)}
=- C_F \frac{C_A}{2}\frac{1+\left(1+\lambda_1^- - y \right)\left(1-y\right)}{1+(1-y)^2}
\frac{y - \frac{\lambda_1^-}{2}}{y - \lambda_1^-}
\frac{ \mathbf{l}_{ \perp} \cdot \left(\mathbf{l}_{ \perp} - \mathbf{k}_{1 \perp}\right)+\frac{y^2(y-\lambda_1^-)^2 M^2}{1+(1-y)(1+\lambda_1^- -y)}}{\left[l_\perp^2 + y^2 M^2\right]\left[\left(\mathbf{l}_{ \perp} -\mathbf{k}_{1 \perp} \right)^2+(y-\lambda_1^-)^2 M^2\right] } .
\end{eqnarray}

\begin{figure}[thb]
\centering
\includegraphics[width=0.99\linewidth]{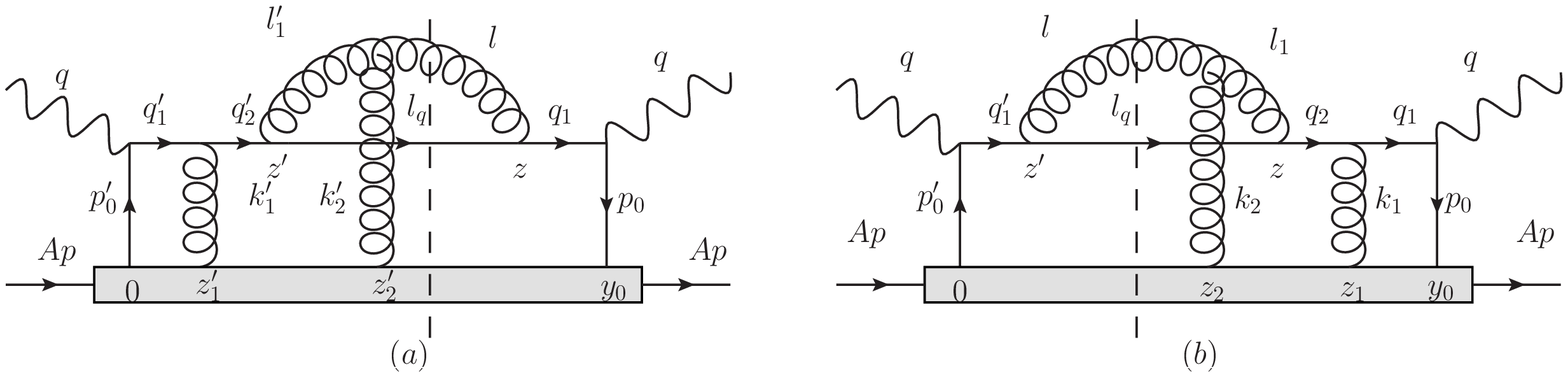}
	\caption{Two non-central-cut diagrams: two rescatterings (with one on the quark before gluon emission and one on the radiated gluon) in the amplitude and zero rescattering in the complex conjugate (a) or vice versa (b).}
	\label{heavy9}
\end{figure}
The phase factor for Figure~\ref{heavy9} reads:
\begin{eqnarray}
S_{(\ref{heavy9},a)}&=&\frac{1}{2} e^{i (\zeta_{M1}'-\zeta_{M0}') p^+ z_1'^-}e^{i (\tilde{x}_L (1-y) - y\, x_M-\lambda_{D1}')p^+ z_2'^-}\left(e^{i \chi_{M11}' p^+ z_1'^-} - e^{i \chi_{M11}' p^+  z_2'^-}\right)
\approx0,
\nonumber\\
S_{(\ref{heavy9},b)}&=&\frac{1}{2} e^{- i \zeta_{M0} p^+y_0^-}e^{- i (\zeta_{M1}-\zeta_{M0}) p^+ z_1^-}e^{- i (\tilde{x}_L (1-y) - y\, x_M-\lambda_{D1})p^+ z_2^-}\left(e^{-i \chi_{M11} p^+ z_1^-} - e^{-i \chi_{M11} p^+  z_2^-}\right)
\approx0,
\nonumber\\
S_{(\ref{heavy9})}&=&S_{(\ref{heavy9},a)}\,\,+\,\,S_{(\ref{heavy9},b)}\,\,\approx\,\, 0,
\end{eqnarray}
were $\chi_{M11}=\eta_{M0} +\lambda_{D1}- \zeta_{M1}$. The matrix element for Figure~\ref{heavy9} reads:
\begin{eqnarray}
\hspace{-5mm}
\tilde{T}_{(\ref{heavy9},a)}=\tilde{T}_{(\ref{heavy9},b)}
&=& C_F \frac{C_A}{2}\frac{1+\left(1-\frac{y+\lambda_1^-}{1+\lambda_1^-} \right)\left(1-y\right)}{1+(1-y)^2}
\frac{y-\frac{\lambda_1^-}{2}}{y - \lambda_1^-}
\frac{\mathbf{l}_{\perp}\cdot \left(\mathbf{l}_{\perp} - \frac{y+\lambda_1^-}{1+\lambda_1^-}\mathbf{k}_{1 \perp}\right)+\frac{y^2\left(\frac{y+\lambda_1^-}{1+\lambda_1^-}\right)^2M^2}{1+(1-y)
\left(1-\frac{y+\lambda_1^-}{1+\lambda_1^-}\right)}}{\left[l_\perp^2 + y^2 M^2\right]\left[\left(\mathbf{l}_{\perp} - \frac{y+\lambda_1^-}{1+\lambda_1^-}\mathbf{k}_{1 \perp} \right)^2+\left(\frac{y+\lambda_1^-}{1+\lambda_1^-}\right)^2 M^2\right]}.
\end{eqnarray}

\begin{figure}[thb]
\centering
\includegraphics[width=0.99\linewidth]{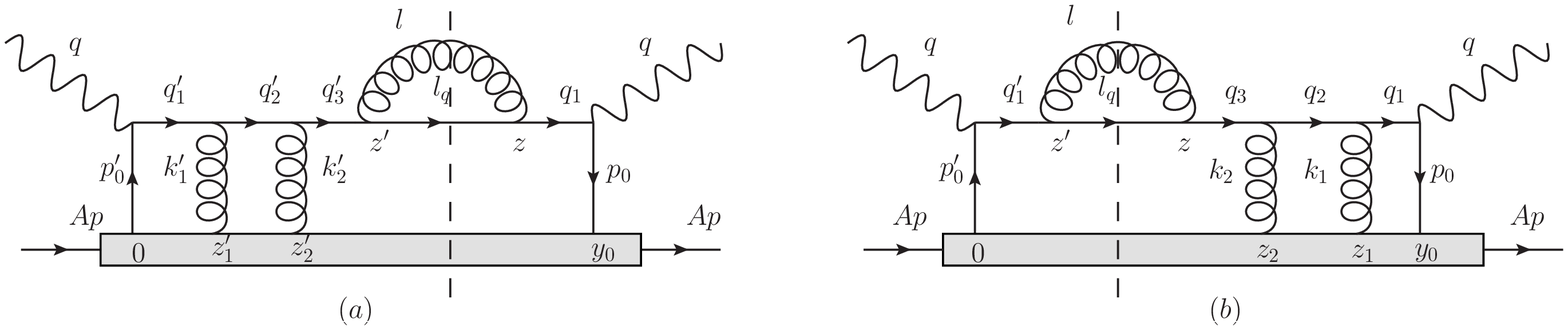}
	\caption{Two non-central-cut diagrams: two rescatterings on the quark before gluon emission in the amplitude and zero rescattering in the complex conjugate (a) or vice versa (b).}
	\label{heavy10}
\end{figure}
The phase factor for Figure~\ref{heavy10} reads:
\begin{eqnarray}
S_{(\ref{heavy10},a)}&=&-\frac{1}{2}e^{i (\zeta_{M1}'-\zeta_{M0}') p^+ z_1'^-}e^{-i (\zeta_{M1}'-\zeta_{M0}') p^+ z_2'^-}e^{-i \tilde{x}_L p^+ z_2'^-}\;\approx\;
-\frac{1}{2}e^{i \tilde{x}_L p^+ Z_1^-},
\nonumber\\
 S_{(\ref{heavy10},b)}&=&-\frac{1}{2}e^{-i\zeta_{M0}p^+ y_0^-}e^{-i (\zeta_{M1}-\zeta_{M0}) p^+ z_1^-}e^{i (\zeta_{M1}-\zeta_{M0}) p^+ z_2^-}e^{-i \tilde{x}_L p^+ z_2^-}\;\approx\;
 -\frac{1}{2}e^{i \tilde{x}_L p^+ Z_1^-},
\nonumber\\
S_{(\ref{heavy10})}&=& S_{(\ref{heavy10},a)}\,\,+\,\, S_{(\ref{heavy10},b)}\,\,\approx\,\,
-\cos(\tilde{x}_L p^+ Z_1^-).
\end{eqnarray}
The matrix element for Figure~\ref{heavy10} reads:
\begin{eqnarray}
\tilde{T}_{(\ref{heavy10},a)}=\delta\bar{ T}_{(\ref{heavy10},b)}
=C_F^2\frac{l_\perp^2+\frac{y^4 M^2}{1+(1-y)^2}}{[l_\perp^2 + y^2 M^2]^2}.
\end{eqnarray}

\begin{figure}[thb]
\centering
\includegraphics[width=0.99\linewidth]{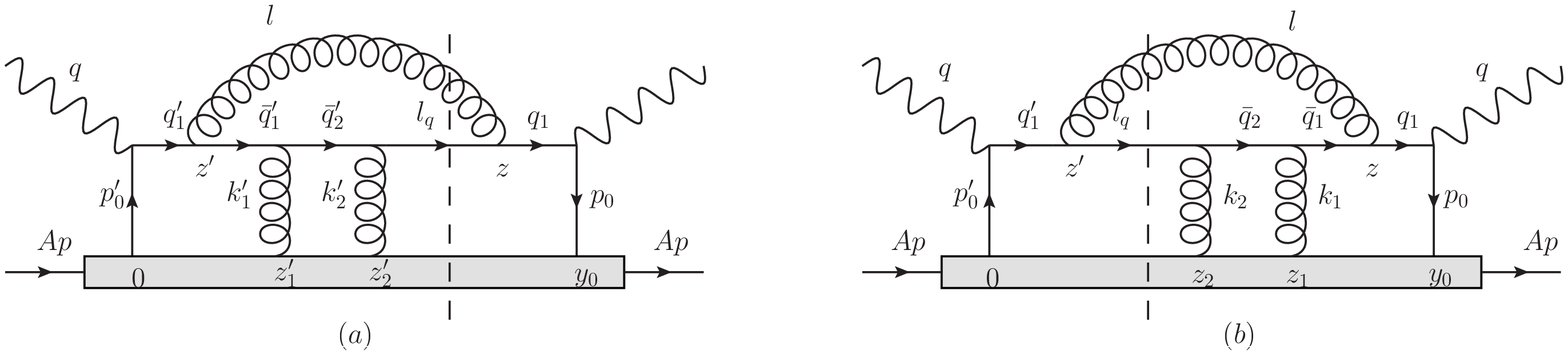}
	\caption{Two non-central-cut diagrams: two rescatterings on the quark after gluon emission in the amplitude and zero rescattering in the complex conjugate (a) or vice versa (b).}
	\label{heavy11}
\end{figure}
The phase factor for Figure~\ref{heavy11} reads:
\begin{eqnarray}
S_{(\ref{heavy11},a)}&=&-\frac{1}{2}e^{i (\eta_{M1}'-\eta_{M0}') p^+ z_1'^-}e^{ -i (\eta_{M1}'-\eta_{M0}') p^+ z_2'^-}(1-e^{i \tilde{x}_L p^+ z_1'^-})
\approx
-\frac{1}{2}(1-e^{i \tilde{x}_L p^+ Z_1^-}),
\nonumber\\
S_{(\ref{heavy11},b)}&=&-\frac{1}{2} e^{-i \zeta_{M0} p^+ y_0^-}e^{- i (\eta_{M1}-\eta_{M0}) p^+ z_1^-}e^{ i (\eta_{M1}-\eta_{M0}) p^+ z_2^-}(e^{-i \tilde{x}_L p^+ y_0^-}-e^{-i \tilde{x}_L p^+ z_1^-})
\approx
-\frac{1}{2}(1-e^{-i \tilde{x}_L p^+ Z_1^-}),
\nonumber\\
S_{(\ref{heavy11})}&=&S_{(\ref{heavy11},a)}\,\,+\,\,S_{(\ref{heavy11},b)}
\,\,\approx\,\,
\cos(\tilde{x}_L p^+ Z_1^-)-1.
\end{eqnarray}
The matrix element for Figure~\ref{heavy11} reads:
\begin{eqnarray}
\tilde{T}_{(\ref{heavy11},a)}=\tilde{T}_{(\ref{heavy11},b)}
=C_F^2\frac{l_\perp^2+\frac{y^4 M^2}{1+(1-y)^2}}{[l_\perp^2 + y^2 M^2]^2}.
\end{eqnarray}

\begin{figure}[thb]
\centering
\includegraphics[width=0.99\linewidth]{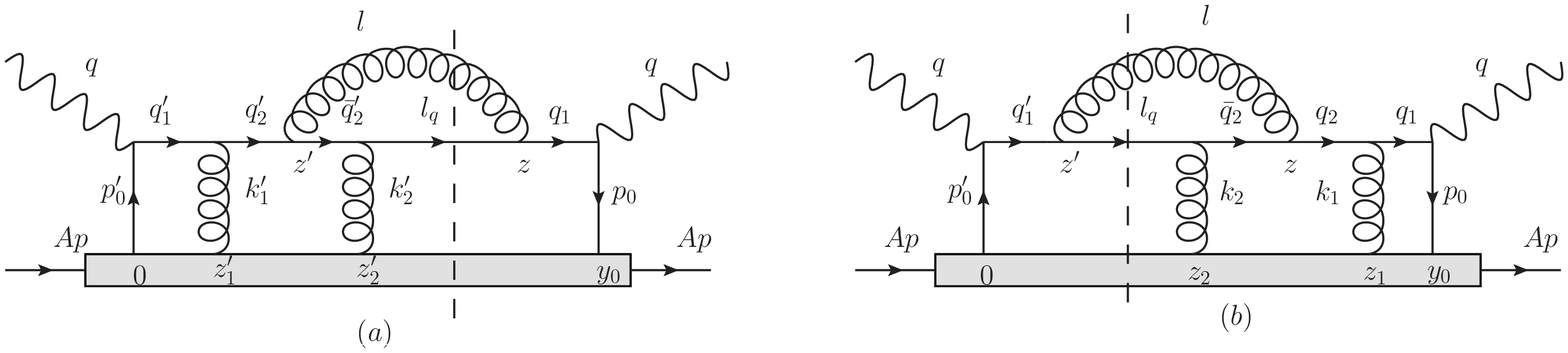}
	\caption{Two non-central-cut diagrams: two rescatterings (with one on the quark before gluon emission and one on the quark after gluon emission) in the amplitude and zero rescattering in the complex conjugate (a) or vice versa (b).}
	\label{heavy12}
\end{figure}
The phase factor for Figure~\ref{heavy12} reads:
\begin{eqnarray}
 S_{(\ref{heavy12},a)}\,\,&=&\,\,-\frac{1}{2}e^{-i \eta_{M0}p^+ y_0^-}e^{i \eta_{M1}' p^+ z_1'^-}  e^{i\eta_{M2}' p^+ z_2'^-}\left(e^{i \tilde{x}_L p^+ z_1'^- }-e^{i \tilde{x}_L p^+ z_2'^- }\right)
\approx \,0,
\nonumber\\
S_{(\ref{heavy12},b)}\,\,&=&\,\,-\frac{1}{2}e^{-i \zeta_{M0} p^+ y_0^-}e^{-i \eta_{M1} p^+ z_1^-}  e^{-i\eta_{M2} p^+ z_2^-}\left(e^{-i \tilde{x}_L p^+ z_1^- }-e^{-i \tilde{x}_L p^+ z_2^- }\right)
\approx0,
\nonumber\\ S_{(\ref{heavy12})}\,\,&=&\,\,S_{(\ref{heavy12},a)}\,\,+\,\,S_{(\ref{heavy12},b)}\,\,\approx\,\, 0,
\end{eqnarray}
where
\begin{eqnarray}
\eta_{M2}=\frac{l_\perp^2+M^2}{2 p^+ q^-(1-y)}=\eta_{M0}.
\end{eqnarray}
The matrix element for Figure~\ref{heavy12} reads:
\begin{eqnarray}
\hspace{-5mm}
\tilde{T}_{(\ref{heavy12},a)}=\tilde{T}_{(\ref{heavy12},b)}
=C_F\left(C_F-\frac{C_A}{2}\right)\frac{1+(1-y)\left(1-\frac{y}{1+\lambda_1^-}\right)}{1+(1-y)^2}
\frac{\mathbf{l}_{\perp} \cdot \left(\mathbf{l}_{\perp} - \frac{y}{1+\lambda_1^-} \mathbf{k}_{1\perp}\right)+\frac{y^2\left(\frac{y}{1+\lambda_1^-}\right)^2M^2}
{1+(1-y)\left(1-\frac{y}{1+\lambda_1^-}\right)}}{\left[l_\perp^2+y^2 M^2\right] \left[\left(\mathbf{l}_{\perp} - \frac{y}{1+\lambda_1^-} \mathbf{k}_{1\perp}\right)^2+\left(\frac{y}{1+\lambda_1^-}\right)^2 M^2\right]}.
\end{eqnarray}

\section*{APPENDIX B}

In this Appendix, we provide more details on the origin of the distribution function $\mathcal{D}(k_1^-, \mathbf{k}_{1\perp})$. We consider the DIS process in which the produced light quark jet rescatters with the medium constituent, as shown in Figure~\ref{el_cc}.

\begin{figure}[thb]
\centering
\includegraphics[width=0.49\linewidth]{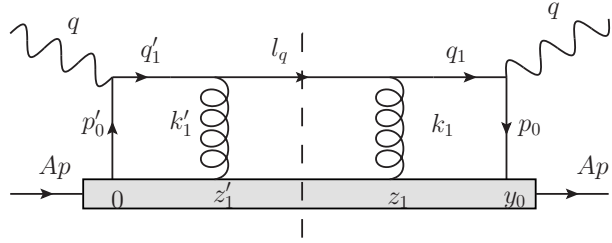}
\caption{A central-cut diagram: the produced light quark experiences one rescattering with the medium in both the amplitude and the complex conjugate.
}
\label{el_cc}
\end{figure}

Here we focus on the central-cut diagram in Figure~\ref{el_cc} where the produced hard quark changes its momentum via rescattering with the medium constituents.
The corresponding non-central-cut diagrams, with zero rescattering in either the amplitude or the complex conjugate, do not contribute to the change of the light quark momentum (they are however important to ensure the unitary).
The hadronic tensor for Figure~\ref{el_cc} can be expressed as follows:
\begin{eqnarray}
W_{\rm el}^{A\mu\nu} &=& \frac{1}{N_c} \int \frac{d^4l_q}{(2\pi)^4} (2\pi)\delta(l_q^2) \int d^4y_0 e^{i q\cdot y_0}
 \int d^4 z_1 \int d^4 z_1'\int \frac{d^4q_1}{(2\pi)^4} \int \frac{d^4q_1'}{(2\pi)^4}e^{-iq_1\cdot (y_{0} - z_1)} e^{-iq_1'\cdot (z_1'-y_0')} e^{-il_q\cdot (z_{1} - {z}_{1}')}
\nonumber\\&\times&
|V_{i j}|^2 \langle A | \bar{\psi}_i(y_0) \gamma^\mu (1- \gamma^5) \frac{- i \slashed{q}_1}{q_1^2 - i\epsilon}
(- i g \gamma_{\alpha} T^{a_1}) A_{a_1}^{\alpha}(z_1) \slashed{l}_q
(i g \gamma_{\alpha'}T^{a_1'}) A_{a_1'}^{\alpha'}(z_1')
\frac{i \slashed{q}_1'}{q_1'^2 + i \epsilon} (1 + \gamma^5) \gamma^\nu \psi_i(0) |A\rangle. \;\;\;\;\;
\end{eqnarray}
From momentum conservations in Figure~\ref{el_cc}, we have the following momentum for various momenta,
\begin{eqnarray}
p_0 = q_1 - q,\, k_1 = l_q - q_1,\, p_0' = q_1' - q,\, k_1' = l_q - q_1'.
\end{eqnarray}
To simplify the hadronic tensor, we may re-introduce the momentum variable $p_0 = l_q - q - k_1$ and change the integration variables $q_1 \to k_1$ and $q_1' \to k_1'$.  The phase factor can be expressed as: $e^{-ip_0\cdot y_0} e^{-ik_1\cdot z_1} e^{ik_1'\cdot z_1'}$.
Following the standard procedure in Sec. III, we first look at the internal quark propagator and the external quark line,
\begin{eqnarray}
&&q_{1}^2 = (q + p_0)^2 = 2p^+ q^- (1+x_0^-)\left[ -{x}_B + x_0 - {\zeta}_{0} \right],
\\
&&l_q^2 = (q_1 +k_1)^2= (q + p_0 +k_1)^2 = 2p^+ q^- (1 +x_0^- + \lambda_1^-) \left[ -{x}_B + x_0 + \lambda_1 - {\zeta}_{01} \right],
\nonumber
\end{eqnarray}
where the following momentum fractions are defined,
\begin{eqnarray}
x_0 = \frac{p_0^+}{p^+},\;\lambda_1 = \frac{k_1^+}{p^+},\;\;
x_0^- = \frac{p_0^-}{q^-},\;\lambda_1^- = \frac{k_1^-}{q^-},\;\;
\zeta_{0} = \frac{ \mathbf{p}_{0\perp}^2}{2p^+q^-(1 + x_0^-)}, \;\;\;
\zeta_{01} = \frac{ (\mathbf{p}_{0\perp}+\mathbf{k}_{1\perp})^2}{2p^+q^-(1 + x_0^- +\lambda_1^-)} .
\end{eqnarray}
Similarly, we may also define the momentum fractions $x_0'$, $\lambda_1'$, $x_0'^-$, $\lambda_1'^-$, $\zeta_0'$ and $\zeta_{01}'$.
In limit of very high energy, one can ignore the transverse ($\perp$) components of the quark and gluon field operators and only keep the forward (+) component of the scattered gluon field. By factorizing out one-nucleon state from the nucleus state, the hadronic tensor may be written as:
\begin{eqnarray}
W_{\rm el}^{A\mu\nu} &=& g^2 \int {d^3 \mathbf{l}_q} \int dy_0^- \int d z_1^- \int d z _1'^- \int d^3\mathbf{y}_0 \int d^3\mathbf{z}_1\int d^3\mathbf{z}_1'
\nonumber \\&\times&
\int \frac{d x_0}{2\pi} \int\frac{d \lambda_1}{2\pi} \int\frac{d \lambda_1'}{2\pi}
\frac{1}{(2q^-)^3} \frac{1}{1+x_0^-} \frac{1}{1+x_0'^-} \frac{1}{1+x_0^-+\lambda_1^-}
\nonumber \\&\times&
\frac{e^{-i x_0 p^+ y_0^-}e^{-i \lambda_1 p^+ z_1^-}}{-{x}_B + x_0 - {\zeta}_{0}-i \epsilon}\frac{e^{i \lambda_1' p^+ z_1'^-}}{-{x}_B + x_0' - {\zeta}_{0}'+i \epsilon} (2\pi) \delta(-{x}_B + x_0 + \lambda_1 - \zeta_{01})
\nonumber \\&\times&
\int \frac{d^3\mathbf{p}_0}{(2\pi)^3} \int \frac{d^3\mathbf{k}_1}{(2\pi)^3} \int \frac{d^3\mathbf{k}_1'}{(2\pi)^3}
e^{-i \mathbf{p}_0 \cdot \mathbf{{y}}_0}e^{-i \mathbf{{k}}_1 \cdot \mathbf{z}_1 }e^{i \mathbf{{k}}_1' \cdot \mathbf{z}_1'}
|V_{i j}|^2 A C_p^A \langle p | \bar{\psi}(y_0^-) \frac{\gamma^+}{2} \psi(0) |p\rangle \langle A| A_{a_1}^+(z_1) A_{a_1'}^+(z_1')|A\rangle
\nonumber \\ &\times&
\frac{1}{4 p^+ q^-}{\rm Tr} \left[\slashed{p} \gamma^\mu(1- \gamma^5) \{\slashed{q} + x_B \slashed{p}\} (1+ \gamma^5)\gamma^\nu \right] \frac{1}{N_c} {\rm Tr}\left[T^{a_1} T^{a_1'}\right]
{\rm Tr}\left[ \frac{\gamma^-}{2} \slashed{q}_1 \gamma^- \slashed{l}_q \gamma^- \slashed{q}_1' \right] \delta^3 \left(\mathbf{l}_q-\mathbf{q}-\mathbf{k}_1 \right).
\end{eqnarray}
The hard trace part can be evaluated as:
\begin{eqnarray}
{\rm Tr}\left[ \frac{\gamma^-}{2} \slashed{q}_1 \gamma^- \slashed{l}_q \gamma^- \slashed{q}_1' \right]
= 8( q^-)^3 (1+x_0^-)(1+x_0'^-)(1+x_0^-+\lambda_1^-).
\end{eqnarray}
The integrations over the momentum fractions $x_0$, $\lambda_1$ and $\lambda_1'$ can be performed as follows. One may first perform the integration over $\lambda_1$ by using the $\delta$ function for the on-shell condition of the final quark. Then the integrations over $x_0$ may be carried out with the technique of contour integration,
\begin{eqnarray}
& &\int \frac{dx_{0}}{2\pi} \frac{e^{-ix_0 p^+ (y_0^- -z_1^-)}}{-{x}_B + x_0 - \zeta_{0} - i\epsilon}
= i \theta(z_1^--y_0^-) e^{-i ( {x}_B + \zeta_{0} )p^+ (y_0^- - z_1^-)}.
\end{eqnarray}
By shifting the integration variable from $\lambda_1'$ to $x_0'= x_0 + \lambda_1 - \lambda_1'$, one may carry out the integration over $x_0'$ analogously.
Then we change the coordinate variables $(z_1, z_1')$ to their mean value $Z_1=(z_1 + z'_1)/2$ and their difference $\delta z_1= z_1 - z'_1$.
After some similar steps as performed in Sec. III, we finally obtain the differential hadronic tensor as follows:
\begin{eqnarray}
\frac{dW_{\rm el}^{A\mu\nu}}{d^3 \mathbf{l}_q} = W_0^{A\mu\nu}
\int d Z_1^- \int d\delta z_1^-
 \int d^3 \delta \mathbf{z}_1\int \frac{d^3\mathbf{k}_1}{(2\pi)^3}e^{-i \mathbf{{k}}_1 \cdot {\delta\mathbf{ z}}_1} \left(g^2\frac{C_F C_2(R)}{N_c^2-1}\right) \langle A| A^+(\delta z_1) A^+(0)|A\rangle\delta^3 \left( \mathbf{l}_q-\mathbf{q}-\mathbf{k}_1 \right). \ \ \ \
\end{eqnarray}
where $W_0^{A\mu\nu}$ is the hadronic tensor without any scattering or radiation:
\begin{eqnarray}
W_0^{A\mu\nu} = |V_{ij}|^2 A C_p^A (2\pi) f_i(x_B) \frac{1}{4p^+ q^-} {\rm Tr}[ \slashed{p} \gamma^\mu (1- \gamma_5)(\slashed{q} + x_B \slashed{p}) (1+\gamma_5) \gamma^\nu].
\end{eqnarray}
Now the distribution function $\mathcal{D}(\,k_1^-,\, \mathbf{k}_{1\, \perp})$ can be defined:
\begin{eqnarray}
\mathcal{D}(\,k_1^-,\, \mathbf{k}_{1\, \perp}) = \int d \delta z_1^- \int d^3 \delta \mathbf{z}_{1} e^{-i \mathbf{k}_{1} \cdot \delta \mathbf{z}_{1}}\left( g^2 \frac{C_F\,C_2(R)}{N_c^2-1}\right)\langle A | A^+(\delta z_1^-, \delta \mathbf{ z}_{1}) A^+(0)  |A\rangle.\;\;\;\;\;\;\;\;\;
\end{eqnarray}
Using the distribution function $\mathcal{D}(\,k_1^-,\, \mathbf{k}_{1\, \perp})$, the differential hadronic tensor for Figure~\ref{el_cc} may be written as:
\begin{eqnarray}
\frac{dW_{\rm el}^{A\mu\nu}}{d^3 \mathbf{l}_q} 
= W_0^{A\mu\nu}
\int dZ_1^- \int \frac{dk_1^- d^2\mathbf{k}_{1\perp}}{(2\pi)^3} \mathcal{D}(k_1^-, \mathbf{k}_{1\perp}) \delta^3(\mathbf{l}_q - \mathbf{q} - \mathbf{k}_1).
\end{eqnarray}
The above equation means that the distribution function $\mathcal{D}(k_1^-,\, \mathbf{k}_{1\, \perp})$ may be understood as, up to a constant factor, the differential elastic scattering rate ${dP_{\rm el}}/{dk_1^- d^2\mathbf{k}_{1\perp} dZ_1^-}$ between a light quark and medium constituents.

\end{widetext}

\bibliographystyle{plain}
\bibliographystyle{h-physrev5}
\bibliography{refs_GYQ}

\begin{thebibliography}{10}

\bibitem{Wang:1991xy}
X.-N. Wang and M.~Gyulassy,
\newblock Phys.Rev.Lett. {\bf 68}, 1480 (1992).

\bibitem{Qin:2015srf}
G.-Y. Qin and X.-N. Wang,
\newblock Int. J. Mod. Phys. {\bf E24}, 1530014 (2015), arXiv:1511.00790.

\bibitem{Blaizot:2015lma}
J.-P. Blaizot and Y.~Mehtar-Tani,
\newblock Int. J. Mod. Phys. {\bf E24}, 1530012 (2015), arXiv:1503.05958.

\bibitem{Majumder:2010qh}
A.~Majumder and M.~Van~Leeuwen,
\newblock Prog.Part.Nucl.Phys. {\bf A66}, 41 (2011), arXiv:1002.2206.

\bibitem{Abelev:2012hxa}
ALICE, B.~Abelev {\em et~al.},
\newblock Phys. Lett. {\bf B720}, 52 (2013), arXiv:1208.2711.

\bibitem{Aad:2015wga}
ATLAS, G.~Aad {\em et~al.},
\newblock JHEP {\bf 09}, 050 (2015), arXiv:1504.04337.

\bibitem{CMS:2012aa}
CMS, S.~Chatrchyan {\em et~al.},
\newblock Eur. Phys. J. {\bf C72}, 1945 (2012), arXiv:1202.2554.

\bibitem{Burke:2013yra}
JET, K.~M. Burke {\em et~al.},
\newblock Phys. Rev. {\bf C90}, 014909 (2014), arXiv:1312.5003.

\bibitem{Xu:2014tda}
J.~Xu, J.~Liao, and M.~Gyulassy,
\newblock Chin. Phys. Lett. {\bf 32}, 092501 (2015), arXiv:1411.3673.

\bibitem{Chien:2015vja}
Y.-T. Chien, A.~Emerman, Z.-B. Kang, G.~Ovanesyan, and I.~Vitev,
\newblock Phys. Rev. {\bf D93}, 074030 (2016), arXiv:1509.02936.

\bibitem{Andres:2016iys}
C.~Andr¨¦s, N.~Armesto, M.~Luzum, C.~A. Salgado, and P.~Zurita,
\newblock Eur. Phys. J. {\bf C76}, 475 (2016), arXiv:1606.04837.

\bibitem{Cao:2017hhk}
S.~Cao, T.~Luo, G.-Y. Qin, and X.-N. Wang,
\newblock Phys. Lett. {\bf B777}, 255 (2018), arXiv:1703.00822.

\bibitem{Zigic:2018ovr}
D.~Zigic, I.~Salom, J.~Auvinen, M.~Djordjevic, and M.~Djordjevic,
\newblock Phys. Lett. {\bf B791}, 236 (2019), arXiv:1805.04786.

\bibitem{Adam:2015ewa}
ALICE, J.~Adam {\em et~al.},
\newblock Phys. Lett. {\bf B746}, 1 (2015), arXiv:1502.01689.

\bibitem{Aad:2014bxa}
ATLAS, G.~Aad {\em et~al.},
\newblock Phys. Rev. Lett. {\bf 114}, 072302 (2015), arXiv:1411.2357.

\bibitem{Khachatryan:2016jfl}
CMS, V.~Khachatryan {\em et~al.},
\newblock Phys. Rev. {\bf C96}, 015202 (2017), arXiv:1609.05383.

\bibitem{Qin:2010mn}
G.-Y. Qin and B.~Muller,
\newblock Phys. Rev. Lett. {\bf 106}, 162302 (2011), arXiv:1012.5280,
\newblock [Erratum: Phys. Rev. Lett.108,189904(2012)].

\bibitem{Young:2011qx}
C.~Young, B.~Schenke, S.~Jeon, and C.~Gale,
\newblock Phys.Rev. {\bf C84}, 024907 (2011), arXiv:1103.5769.

\bibitem{Dai:2012am}
W.~Dai, I.~Vitev, and B.-W. Zhang,
\newblock Phys. Rev. Lett. {\bf 110}, 142001 (2013), arXiv:1207.5177.

\bibitem{Wang:2013cia}
X.-N. Wang and Y.~Zhu,
\newblock Phys. Rev. Lett. {\bf 111}, 062301 (2013), arXiv:1302.5874.

\bibitem{Blaizot:2013hx}
J.-P. Blaizot, E.~Iancu, and Y.~Mehtar-Tani,
\newblock Phys.Rev.Lett. {\bf 111}, 052001 (2013), arXiv:1301.6102.

\bibitem{Mehtar-Tani:2014yea}
Y.~Mehtar-Tani and K.~Tywoniuk,
\newblock Phys. Lett. {\bf B744}, 284 (2015), arXiv:1401.8293.

\bibitem{Cao:2017qpx}
S.~Cao and A.~Majumder,
\newblock (2017), arXiv:1712.10055.

\bibitem{He:2018xjv}
Y.~He {\em et~al.},
\newblock (2018), arXiv:1809.02525.

\bibitem{Aad:2010bu}
Atlas Collaboration, G.~Aad {\em et~al.},
\newblock Phys.Rev.Lett. {\bf 105}, 252303 (2010), arXiv:1011.6182.

\bibitem{Chatrchyan:2012gt}
CMS, S.~Chatrchyan {\em et~al.},
\newblock Phys. Lett. {\bf B718}, 773 (2013), arXiv:1205.0206.

\bibitem{Qin:2009bk}
G.-Y. Qin, J.~Ruppert, C.~Gale, S.~Jeon, and G.~D. Moore,
\newblock Phys.Rev. {\bf C80}, 054909 (2009), arXiv:0906.3280.

\bibitem{Chen:2016vem}
L.~Chen, G.-Y. Qin, S.-Y. Wei, B.-W. Xiao, and H.-Z. Zhang,
\newblock Phys. Lett. {\bf B773}, 672 (2017), arXiv:1607.01932.

\bibitem{Chen:2016cof}
L.~Chen, G.-Y. Qin, S.-Y. Wei, B.-W. Xiao, and H.-Z. Zhang,
\newblock Phys. Lett. {\bf B782}, 773 (2018), arXiv:1612.04202.

\bibitem{Chen:2017zte}
W.~Chen, S.~Cao, T.~Luo, L.-G. Pang, and X.-N. Wang,
\newblock Phys. Lett. {\bf B777}, 86 (2018), arXiv:1704.03648.

\bibitem{Luo:2018pto}
T.~Luo, S.~Cao, Y.~He, and X.-N. Wang,
\newblock Phys. Lett. {\bf B782}, 707 (2018), arXiv:1803.06785.

\bibitem{Zhang:2018urd}
S.-L. Zhang, T.~Luo, X.-N. Wang, and B.-W. Zhang,
\newblock Phys. Rev. {\bf C98}, 021901 (2018), arXiv:1804.11041.

\bibitem{Kang:2018wrs}
Z.-B. Kang, J.~Reiten, I.~Vitev, and B.~Yoon,
\newblock Phys. Rev. {\bf D99}, 034006 (2019), arXiv:1810.10007.

\bibitem{Chatrchyan:2013kwa}
CMS Collaboration, S.~Chatrchyan {\em et~al.},
\newblock Phys.Lett. {\bf B730}, 243 (2014), arXiv:1310.0878.

\bibitem{Aad:2014wha}
ATLAS, G.~Aad {\em et~al.},
\newblock Phys. Lett. {\bf B739}, 320 (2014), arXiv:1406.2979.

\bibitem{Chang:2016gjp}
N.-B. Chang and G.-Y. Qin,
\newblock Phys. Rev. {\bf C94}, 024902 (2016), arXiv:1603.01920.

\bibitem{Casalderrey-Solana:2016jvj}
J.~Casalderrey-Solana, D.~Gulhan, G.~Milhano, D.~Pablos, and K.~Rajagopal,
\newblock JHEP {\bf 03}, 135 (2017), arXiv:1609.05842.

\bibitem{Tachibana:2017syd}
Y.~Tachibana, N.-B. Chang, and G.-Y. Qin,
\newblock Phys. Rev. {\bf C95}, 044909 (2017), arXiv:1701.07951.

\bibitem{KunnawalkamElayavalli:2017hxo}
R.~Kunnawalkam~Elayavalli and K.~C. Zapp,
\newblock JHEP {\bf 07}, 141 (2017), arXiv:1707.01539.

\bibitem{Brewer:2017fqy}
J.~Brewer, K.~Rajagopal, A.~Sadofyev, and W.~Van Der~Schee,
\newblock JHEP {\bf 02}, 015 (2018), arXiv:1710.03237.

\bibitem{Chien:2018dfn}
Y.-T. Chien and R.~Kunnawalkam~Elayavalli,
\newblock (2018), arXiv:1803.03589.

\bibitem{Bjorken:1982tu}
J.~D. Bjorken,
\newblock FERMILAB-PUB-82-059-THY.

\bibitem{Braaten:1991we}
E.~Braaten and M.~H. Thoma,
\newblock Phys.Rev. {\bf D44}, 2625 (1991).

\bibitem{Djordjevic:2006tw}
M.~Djordjevic,
\newblock Phys. Rev. {\bf C74}, 064907 (2006), arXiv:nucl-th/0603066.

\bibitem{Qin:2007rn}
G.-Y. Qin {\em et~al.},
\newblock Phys. Rev. Lett. {\bf 100}, 072301 (2008), arXiv:0710.0605.

\bibitem{Qin:2012fua}
G.-Y. Qin and A.~Majumder,
\newblock Phys.Rev. {\bf C87}, 024909 (2013), arXiv:1205.5741.

\bibitem{Baier:1996kr}
R.~Baier, Y.~L. Dokshitzer, A.~H. Mueller, S.~Peigne, and D.~Schiff,
\newblock Nucl.Phys. {\bf B483}, 291 (1997), arXiv:hep-ph/9607355.

\bibitem{Baier:1996sk}
R.~Baier, Y.~L. Dokshitzer, A.~H. Mueller, S.~Peigne, and D.~Schiff,
\newblock Nucl.Phys. {\bf B484}, 265 (1997), arXiv:hep-ph/9608322.

\bibitem{Baier:1998kq}
R.~Baier, Y.~L. Dokshitzer, A.~H. Mueller, and D.~Schiff,
\newblock Nucl. Phys. {\bf B531}, 403 (1998), arXiv:hep-ph/9804212.

\bibitem{Zakharov:1996fv}
B.~Zakharov,
\newblock JETP Lett. {\bf 63}, 952 (1996), arXiv:hep-ph/9607440.

\bibitem{Zakharov:1997uu}
B.~G. Zakharov,
\newblock JETP Lett. {\bf 65}, 615 (1997), arXiv:hep-ph/9704255.

\bibitem{Gyulassy:1999zd}
M.~Gyulassy, P.~Levai, and I.~Vitev,
\newblock Nucl.Phys. {\bf B571}, 197 (2000), arXiv:hep-ph/9907461.

\bibitem{Gyulassy:2000er}
M.~Gyulassy, P.~Levai, and I.~Vitev,
\newblock Nucl.Phys. {\bf B594}, 371 (2001), arXiv:nucl-th/0006010.

\bibitem{Djordjevic:2003zk}
M.~Djordjevic and M.~Gyulassy,
\newblock Nucl. Phys. {\bf A733}, 265 (2004), arXiv:nucl-th/0310076.

\bibitem{Blagojevic:2018nve}
B.~Blagojevic, M.~Djordjevic, and M.~Djordjevic,
\newblock Phys. Rev. {\bf C99}, 024901 (2019), arXiv:1804.07593.

\bibitem{Sievert:2018imd}
M.~D. Sievert and I.~Vitev,
\newblock Phys. Rev. {\bf D98}, 094010 (2018), arXiv:1807.03799.

\bibitem{Wiedemann:2000za}
U.~A. Wiedemann,
\newblock Nucl.Phys. {\bf B588}, 303 (2000), arXiv:hep-ph/0005129.

\bibitem{Wiedemann:2000tf}
U.~A. Wiedemann,
\newblock Nucl.Phys. {\bf A690}, 731 (2001), arXiv:hep-ph/0008241.

\bibitem{Armesto:2003jh}
N.~Armesto, C.~A. Salgado, and U.~A. Wiedemann,
\newblock Phys. Rev. {\bf D69}, 114003 (2004), arXiv:hep-ph/0312106.

\bibitem{Arnold:2001ba}
P.~B. Arnold, G.~D. Moore, and L.~G. Yaffe,
\newblock JHEP {\bf 0111}, 057 (2001), arXiv:hep-ph/0109064.

\bibitem{Arnold:2002ja}
P.~B. Arnold, G.~D. Moore, and L.~G. Yaffe,
\newblock JHEP {\bf 0206}, 030 (2002), arXiv:hep-ph/0204343.

\bibitem{CaronHuot:2010bp}
S.~Caron-Huot and C.~Gale,
\newblock Phys. Rev. {\bf C82}, 064902 (2010), arXiv:1006.2379.

\bibitem{Guo:2000nz}
X.-f. Guo and X.-N. Wang,
\newblock Phys.Rev.Lett. {\bf 85}, 3591 (2000), arXiv:hep-ph/0005044.

\bibitem{Wang:2001ifa}
X.-N. Wang and X.-f. Guo,
\newblock Nucl. Phys. {\bf A696}, 788 (2001), arXiv:hep-ph/0102230.

\bibitem{Zhang:2003wk}
B.-W. Zhang, E.~Wang, and X.-N. Wang,
\newblock Phys. Rev. Lett. {\bf 93}, 072301 (2004), arXiv:nucl-th/0309040.

\bibitem{Majumder:2009ge}
A.~Majumder,
\newblock Phys. Rev. {\bf D85}, 014023 (2012), arXiv:0912.2987.

\bibitem{Armesto:2011ht}
N.~Armesto {\em et~al.},
\newblock Phys. Rev. {\bf C86}, 064904 (2012), arXiv:1106.1106.

\bibitem{Apolinario:2014csa}
L.~Apolinário, N.~Armesto, J.~Milhano, and C.~A. Salgado,
\newblock JHEP {\bf 02}, 119 (2015), arXiv:1407.0599.

\bibitem{Zhang:2018kkn}
L.~Zhang, D.-F. Hou, and G.-Y. Qin,
\newblock Phys. Rev. {\bf C98}, 034913 (2018), arXiv:1804.00470.

\bibitem{ZhangYY}
Y.~Zhang, G.-Y. Qin, and X.-N. Wang,
\newblock to be published.

\bibitem{Gyulassy:2000gk}
M.~Gyulassy, I.~Vitev, and X.~N. Wang,
\newblock Phys. Rev. Lett. {\bf 86}, 2537 (2001), arXiv:nucl-th/0012092.

\bibitem{Aivazis:1993kh}
M.~A.~G. Aivazis, F.~I. Olness, and W.-K. Tung,
\newblock Phys. Rev. {\bf D50}, 3085 (1994), arXiv:hep-ph/9312318.

\bibitem{Altarelli:1977zs} 
G.~Altarelli and G.~Parisi,
\newblock Nucl.\ Phys.\ B {\bf 126}, 298 (1977).

\bibitem{Djordjevic:2007at}
M.~Djordjevic and U.~Heinz,
\newblock Phys. Rev. {\bf C77}, 024905 (2008), arXiv:0705.3439.

\bibitem{Djordjevic:2008iz}
M.~Djordjevic and U.~W. Heinz,
\newblock Phys.Rev.Lett. {\bf 101}, 022302 (2008), arXiv:0802.1230.

\end{thebibliography}
\end{document}